\documentclass[11pt,letterpaper]{article}
\usepackage{amsmath,amsfonts,amssymb}
\usepackage{graphicx}
\usepackage{float}
\usepackage{booktabs,adjustbox}
\usepackage[margin=1in]{geometry}
\usepackage[round]{natbib}
\usepackage{amsthm}
\newtheorem{theorem}{Theorem}
\def\b0{{\bf 0}}
\newcommand{\bl}{\begin{flushleft}}
\newcommand{\el}{\end{flushleft}}
\newcommand{\bc}{\begin{center}}
\newcommand{\ec}{\end{center}}

\begin{document}

\title{Estimating Group Average Treatment Effects in High-Risk Populations with Unobserved Target Membership}

\author{Shubo Zhang\(^{1}\), Yinjun Zhao\(^{1}\), Donglin Zeng\(^{2}\), and Yuanjia Wang\(^{1}\)\\
\(^{1}\)Department of Biostatistics, Mailman School of Public Health, Columbia University\\
\(^{2}\)Department of Biostatistics, School of Public Health, University of Michigan\\
Correspondence: Yuanjia Wang (yw2016@cumc.columbia.edu)}
\date{}

\maketitle

\begin{abstract}
Estimating treatment effects within a prespecified high-risk population is challenging. When risk status is unobserved or the target population is small, analyses restricted to high-risk individuals can be unstable or inefficient. We propose a risk-score-weighted soft-targeting (RSW) estimator for group average treatment effects in high-risk populations. It accommodates two settings: when high-risk status is observed, the membership probability can be estimated within the analysis sample; when unobserved, it can be obtained from a reference model using covariates in the analysis sample. RSW uses these probabilities to represent the target population, learns a doubly robust treatment effect in the full sample, and aggregates a low-dimensional treatment-effect surface over the risk-weighted target distribution. We establish pointwise asymptotic normality and an influence-function-based variance estimator. In simulations with continuous and binary outcomes, RSW had similar bias and lower variance than two hard-targeting alternatives. Sensitivity analyses showed that the gain persisted when treatment-effect sufficiency was violated, although RSW bias increased under strong violations. We applied the method to the All\textit{of}Us study to estimate effects of non-GLP antidiabetic and antihypertensive medication use on mild cognitive impairment (MCI)-free time restricted at 48 months within comorbidity-defined high-risk subgroups. We estimated the risk score in a nonoverlapping internal reference sample and applied it to the outcome-analysis sample. This illustrates how soft targeting can improve precision while retaining an interpretable high-risk target population.
\end{abstract}


\section{Introduction}

Alzheimer's disease (AD) and related dementias (ADRD) are clinically and biologically heterogeneous syndromes, with variation in presentation, pathology, progression, comorbidity, genetic susceptibility, and social context \citep{Duara2022ADHeterogeneity}. Because vascular, metabolic, behavioral, sensory, and social factors may be modifiable across the life course \citep{Livingston2024}, prevention research must address not only whether an intervention works on average, but also who is most likely to benefit.

Average effects in dementia-prevention trials have been mixed, and reviews of multidomain interventions report at most small cognitive benefits with insufficient evidence for preventing incident dementia \citep{Coley2022RCTReview}. FINGER found cognitive benefit in older adults at elevated risk \citep{Ngandu2015FINGER}, whereas MAPT and preDIVA did not detect significant overall benefit on their primary outcomes \citep{Andrieu2017MAPT}. In SPRINT MIND, intensive blood-pressure control reduced mild cognitive impairment (MCI) and the composite of MCI or probable dementia, but not probable dementia alone \citep{Williamson2019}. These results motivate investigation of heterogeneous treatment effects (HTEs), particularly in populations at increased risk of disease progression.

One challenge is how to characterize factors that contribute to treatment-effect heterogeneity. Prior studies have examined modifiers such as APOE genotype, amyloid status, and predicted cognitive risk \citep[e.g.,][]{Delrieu2019MAPTAmyloid,Ghazi2023SPRINTRisk,Solomon2018FINGERAPOE}.  However, one-at-a-time subgroup analyses can be underpowered and unstable when multiple correlated factors generate small strata \citep{Wang2021SubgroupStatistical}. We therefore consider low-dimensional, clinically interpretable subgroups that summarize multivariable profiles spanning socioeconomic factors, behavioral risks, and comorbidity burden. Beyond characterizing how treatment effects vary across such subgroups, it is also important to define the population within which those effects are most relevant.

In AD/ADRD prevention, this target population may be individuals at increased risk of future disease progression. AD pathological processes may precede MCI or dementia by many years \citep{Jack2018NIAAAFramework}, while clinical progression over typical study horizons can be limited. Trials focusing on clinical endpoints can require thousands of participants and years of follow-up \citep{Brookmeyer2019PreventionTrialDesign}. Enriching for participants at higher risk of decline can therefore improve power and reduce required sample size \citep{Soldan2016PreclinicalGroups}. At the same time, restricting estimation to the observed high-risk group can sacrifice precision when that subset is small. Prognostic risk scores provide an opportunity to use information from the whole population while targeting inference to the high-risk population, potentially improving the efficiency of treatment-effect estimation.

Combining these goals, we aim to estimate group average treatment effects \citep[GATE,][]{chernozhukov2025generic,park2024groupwise} within a target high-risk population. Let \(A\in\{0,1\}\) denote treatment, \(Y(a)\) the potential outcome under treatment $a$, \(D\) a disease risk state used to define the target population, \(c\) the high-risk threshold, and \(Z\) a clinically interpretable subgroup variable. The estimand is
\begin{equation}
\psi(z)=E\{Y(1)-Y(0)\mid D>c, Z=z\}.
\label{eq:target}
\end{equation}
Here, $Y$ and $D$ need not represent the same outcome. The outcome $Y$ is the endpoint on which the intervention effect is evaluated, whereas $D$ determines membership in the target population. Thus, $\psi(z)$ describes the treatment effect on $Y$ within subgroup $Z=z$ among individuals belonging to the high-risk population defined by $D$.

We distinguish two settings. In Setting~I, $D$ and hence high-risk membership are observed in the analysis cohort. In Setting~II, the target population is defined by a latent or later-ascertained disease state $D$ that is unavailable in the analysis cohort. This situation arises naturally in AD/ADRD research. For example, $D$ may represent underlying AD pathology or biomarker positivity, such as amyloid or tau positivity, whereas $Y$ may be an earlier cognitive outcome such as MCI or an outcome affected through vascular or metabolic pathways. Thus, the intervention outcome and the disease state defining the target population may be biologically and temporally distinct. AD pathological status may require amyloid or tau PET imaging, cerebrospinal-fluid biomarkers, or other measurements available only in deeply phenotyped cohorts or subsets \citep{Jack2018NIAAAFramework}. In this setting, a prognostic model developed from external longitudinal or deeply phenotyped data can estimate an individual's probability of target membership, allowing the treatment effect on $Y$ to be estimated within the population defined by $D$ even when $D$ itself is unavailable.

Existing methods for estimating GATEs and HTEs include generic machine-learning summaries and data-adaptive groups in randomized trials \citep{chernozhukov2025generic,imai2025statistical}, semiparametric residualization and orthogonalization \citep{chernozhukov2018double,robinson1988root}, flexible CATE estimators including IPW/AIPW, R-, and DR-learners \citep{kennedy2023towards,wager2018causal}, and inference for groupwise estimands \citep{park2024groupwise}. Related principal-score methods \citep{DingLu2017,FrangakisRubin2002} use covariate-based probabilities of latent stratum membership to identify or estimate effects within latent populations under additional ignorability conditions. Our problem differs in that the target is a prespecified high-risk population, potentially with unobserved membership, and the goal is to estimate subgroup-specific treatment effects within that population while retaining information from the broader analysis cohort.

When $D$ is observed, two natural hard-targeting approaches are to restrict the entire analysis to observed high-risk participants or to estimate treatment and outcome nuisance functions in the full cohort but aggregate treatment-effect estimates only among observed high-risk individuals. Both approaches can become unstable when high-risk subgroups are small, a particular concern for clinical MCI or dementia endpoints \citep{brookmeyer2019design}. More importantly, neither approach is directly applicable when $D$, and hence high-risk membership, is unavailable in the analysis cohort. A hard classification based on an external prediction model is possible, but discards uncertainty in target population membership and may introduce bias.

We therefore propose risk-score-weighted (RSW) soft targeting, which represents the target population by each individual's probability of high-risk membership, learns a doubly robust treatment-effect surface using the full analysis sample, and aggregates this surface over the risk-weighted target distribution. The framework accommodates both settings: when $D$ is observed, the membership probability is estimated within the study; when $D$ is unavailable, it is estimated from an external or reference sample and transported to the analysis cohort. This paper (i) formalizes the high-risk GATE with observed or unobserved target membership; (ii) develops RSW by combining probabilistic target membership with debiased machine-learning estimation of treatment effects; and (iii) establishes asymptotic inference accounting for estimation of the membership score, including a two-sample influence-function formulation when the score is learned externally, and characterizes the potential efficiency gain of soft relative to hard targeting. We evaluate the method in AD/ADRD-motivated simulations and an application to the All\textit{of}Us Research Program \citep{AllofUs2019}, estimating effects of antihypertensive and non-GLP antidiabetic medication use on MCI within a cardiometabolic high-risk population.

\section{Methods}

We describe two settings to estimate \(\psi(z)\) in Equation~\eqref{eq:target}. Setting I (risk status and target membership observed): the observed data are \(O_i=(Y_i,A_i,X_i,Z_i,D_i)\), where \(X_i\) is a vector of potentially high-dimensional baseline covariates, \(Z_i\in\{1,\ldots,G\}\) is the prespecified subgroup, and \(H_i=\mathbb I(D_i>c)\) indicates membership in the high-risk target population. Define the high-risk membership score as
\begin{equation}
W(x,z)=\Pr(H_i=1\mid X_i=x,Z_i=z),\label{eq:wscore}
\end{equation}
and write \(W_i=W(X_i,Z_i)\). When $D_i$ is observed in the study, and thus $H_i$ is available, model \eqref{eq:wscore} can be estimated internally. Setting II (unobserved target membership): $D_i$, and thus the target membership $H_i$, are not available. The observed data are then $O_i=(Y_i,A_i,X_i,Z_i)$, and $W(x,z)$ is computed from an externally reported model \eqref{eq:wscore}. The estimand remains the treatment effect among the high-risk population, $\psi(z)$.

The high-risk score \(W(X,Z)\) is analogous to the propensity score for target membership, where it satisfies the balancing property
\[
H_i \perp\!\!\!\perp X_i \mid W_i, Z_i.
\]
Thus, within a subgroup \(Z=z\), conditioning on the scalar score \(W\) balances the distribution of the potentially high-dimensional covariates \(X\) between high- and lower-risk individuals. This property follows from the definition of \(W\) and does not involve the outcome $Y$. The score therefore serves as a low-dimensional representation of target membership.

\subsection{Identification and assumptions}

We use the following assumptions. \textbf{(A1) Consistency:} \(Y_i=A_iY_i(1)+(1-A_i)Y_i(0)\). \textbf{(A2) Unconfoundedness:} for \(a=0,1\), \(Y_i(a)\perp\!\!\!\perp A_i\mid X_i,Z_i\). \textbf{(A3) Treatment positivity:} for \(\pi(x,z,w)=\Pr(A_i=1\mid X_i=x,Z_i=z,W_i=w)\), all \((x,z,w)\) in the support of \((X_i,Z_i,W_i)\), and some \(\epsilon_A>0\), \(\epsilon_A\le\pi(x,z,w)\le1-\epsilon_A\).
\textbf{(A4) Treatment-effect sufficiency and risk-score validity:} for each subgroup \(z\), $P(H_i=1|Z_i=z)>0,$ and
\begin{equation*}
E\{Y_i(1)-Y_i(0)\mid H_i,W_i,Z_i\}=E\{Y_i(1)-Y_i(0)\mid W_i,Z_i\}.
\end{equation*}
For Setting~II, we additionally assume that the externally obtained model for $W(x,z)$ is transportable to the analysis population, such that it represents the same conditional membership probability over the covariate support in the analysis.

Assumptions (A1)--(A3) are the standard causal identification conditions for observational treatment effects \citep{HernanRobins2020,ImbensRubin2015}. Assumption (A4) is the additional condition for the proposed soft-targeting estimator. It states that, within levels of \(Z_i\) and the high-risk membership score \(W_i\), observed high-risk status contains no additional information about the conditional treatment effects.

When the disease state is assessed post-intervention, Assumption (A4) is related to principal-score methods for causal effects in latent strata. Principal stratification defines causal effects within subpopulations characterized by potential values of an intermediate or post-treatment variable \citep{FrangakisRubin2002}, and the principal score is the conditional probability of membership in a latent principal stratum given pretreatment covariates \citep{DingLu2017}. Principal-score methods assume variants of principal ignorability, under which potential outcomes are independent of principal-stratum membership conditional on sufficiently rich baseline covariates. In the case of AD, interventions such as the antihypertensive and antidiabetic therapies considered in the application here are not expected to change the underlying AD pathological state, so that $D(1)=D(0)=D$, which is a simpler case of principal stratification. Our Assumption~(A4) does not require each potential outcome $Y(a)$ to be conditionally independent of $H$ (or $D$), but requires only the treatment-effect contrast $Y(1)-Y(0)$ to be conditionally independent of $H$ given $W$ and $Z$.

Define \(\tau(z,w)=E\{Y_i(1)-Y_i(0)\mid Z_i=z,W_i=w\}\). Because \(E(H_i\mid W_i,Z_i)=W_i\), Assumptions (A1)--(A4) imply (see Appendix~A.1)
\begin{equation}
\psi(z)=\frac{E\{\tau(z,W_i)W_i\mid Z_i=z\}}{E(W_i\mid Z_i=z)}.
\label{eq:psi_weighted}
\end{equation}
Identification of $\psi(z)$ depends on the treatment-effect sufficiency and risk-score validity condition, without direct observation of $H_i$ or $D_i$. Thus, equation~\eqref{eq:psi_weighted} motivates a soft-targeting estimator to learn a treatment-effect surface indexed by \((Z,W)\) and aggregate it over the risk-weighted distribution within each subgroup. In Setting I, this allows individuals not in the realized high-risk group to contribute to estimation through their membership probabilities, improving precision when the high-risk subgroups in the analysis sample are small. In Setting II, this allows estimating $\psi(z)$ without observing high-risk membership $H$.

\subsection{Three-step estimation}

We estimate \(\psi(z)\) in three steps. First, when $W_i$ are known, let \(\mu_a(x,z,w)=E(Y_i\mid A_i=a,X_i=x,Z_i=z,W_i=w)\) for \(a\in\{0,1\}\), and let \(\pi(x,z,w)=\Pr(A_i=1\mid X_i=x,Z_i=z,W_i=w)\). Write \(\eta=(\mu_0,\mu_1,\pi)\) for the nuisance functions.  Following the interactive regression model (IRM) for a binary treatment \citep{chernozhukov2018double,park2024groupwise}, write
\begin{gather*}
    Y_i=\mu_0(X_i,Z_i,W_i)+A_i\gamma(X_i,Z_i,W_i)+U_i,
\quad E(U_i\mid A_i,X_i,Z_i,W_i)=0,\\
A_i=\pi(X_i,Z_i,W_i)+V_i,\quad
E(V_i\mid X_i,Z_i,W_i)=0,
\end{gather*}
where \(\gamma(x,z,w)=\mu_1(x,z,w)-\mu_0(x,z,w)\). The outcome regressions \(\mu_0,\mu_1\) and treatment propensity \(\pi\) are estimated using flexible learners with \((X_i,Z_i,W_i)\) as features. Given nuisance functions \(\eta\), define the doubly robust orthogonal signal \citep{chernozhukov2018double,RobinsRotnitzky1995}
\[
\Gamma_i(\eta)=\mu_{1,i}-\mu_{0,i}+\frac{A_i}{\pi_i}(Y_i-\mu_{1,i})-\frac{1-A_i}{1-\pi_i}(Y_i-\mu_{0,i}),
\]
where \(\mu_{a,i}=\mu_a(X_i,Z_i,W_i)\) and \(\pi_i=\pi(X_i,Z_i,W_i)\).

At the true nuisance functions \(\eta_0\), Assumptions~(A1)--(A3) and iterated expectations imply \(E\{\Gamma_i(\eta_0)\mid Z_i=z,W_i=w\}=\tau(z,w)\). Thus, \(\Gamma_i(\eta_0)\) is an orthogonal pseudo-outcome whose conditional mean given \((Z_i,W_i)\) is the treatment-effect surface of interest.

Second, we model the treatment effect surface as \(\tau(z,w)=b(z,w)^T\beta_0\), where $b(z,w)$ is a prespecified basis function. This includes linear models and, with spline basis, flexible approximations to nonlinear treatment-effect surfaces. We require either correct specification of the treatment-effect surface by \(b(z,w)^\top\beta_0\), or more generally that the population projection preserves the risk-weighted target as specified in condition (B5) in Appendix~A. Following \citet{SemenovaChernozhukov2021}, \(\beta_0=\arg\min_{\beta}E[\{\Gamma_i(\eta_0)-b(Z_i,W_i)^\top\beta\}^2]\) is the best linear projection coefficient.
One useful specification is \(b(z,w)=\{\mathbf I(z)^\top,p(w)^\top\}^\top\), where \(\mathbf I(z)\) contains subgroup indicators and \(p(w)\) is a series basis such as polynomials or splines. We estimate \(\beta_0\) by regressing \(\Gamma_i\) on \(b(Z_i,W_i)\):
\[
\widehat\beta=\left\{\frac{1}{n}\sum_{i=1}^n b(Z_i,W_i)b(Z_i,W_i)^\top\right\}^{-1}\left\{\frac{1}{n}\sum_{i=1}^n b(Z_i,W_i)\Gamma_i\right\}.
\]
The fitted treatment-effect surface is \(\widehat\tau(z,w)=b(z,w)^\top\widehat\beta\).

Third, when $W_i$ are not known, in Setting I, we estimate the high-risk score by \(L_2\)-penalized logistic regression under
\[
W(x,z;\alpha)=\operatorname{expit}\left\{\alpha_0+x^\top\alpha_X+
\sum_{g=2}^G\alpha_{Z,g}\mathbb I(z=g)\right\},
\]
where \(\alpha=(\alpha_0,\alpha_X^\top,\alpha_{Z,2},\ldots,\alpha_{Z,G})^\top\) and $\widehat W_i= W(X_i,Z_i; \widehat\alpha)$. In Setting II, this first step is replaced by the external target-membership model: $\widehat W_i= W^*(X_i,Z_i)$, where the model $W^*$ may be trained in an independent reference cohort or available through the literature. Replacing $W_i$ by $\widehat W_i$ in the first two steps will give an estimator of GATE.

\subsection{Cross-fitted estimators}
We implement the procedure using a random \(K\)-fold partition \(\mathcal I_1,\ldots,\mathcal I_K\) of \([n]=\{1,\ldots,n\}\), with \(\mathcal I_k^c=[n]\setminus\mathcal I_k\). For each fold, we estimate \(W(x,z;\alpha)\), \(\mu_a(x,z,w)\), and \(\pi(x,z,w)\) on \(\mathcal I_k^c\), using \((X_i,Z_i,\widehat W_i)\) as features for the latter two models. For \(i\in\mathcal I_k\), set \(\widehat\mu_{a,i}=\widehat\mu_a^{(-k)}(X_i,Z_i,\widehat W_i)\), \(a=0,1\), and \(\widehat\pi_i=\widehat\pi^{(-k)}(X_i,Z_i,\widehat W_i)\). For Setting~II, the same cross-fitting is used for the treatment and outcome nuisance functions, but the fold-specific internal estimation of $W$ is omitted. Each held-out subject instead receives the externally generated $\widehat W_i$.

The cross-fitted doubly robust orthogonal signal is
\[
\widehat\Gamma_i
:=
\Gamma_i(\widehat\eta^{(-k)})
=
\widehat\mu_{1,i}-\widehat\mu_{0,i}
+
\frac{A_i}{\widehat\pi_i}\{Y_i-\widehat\mu_{1,i}\}
-
\frac{1-A_i}{1-\widehat\pi_i}\{Y_i-\widehat\mu_{0,i}\},
\qquad i\in\mathcal I_k.
\]
Substituting \(\widehat W_i\) and \(\widehat\Gamma_i\) into the projection above yields the cross-fitted coefficient \(\widehat\beta\) and surface \(\widehat\tau(z,w)=b(z,w)^\top\widehat\beta\).
The proposed risk-score-weighted soft-targeting (RSW) estimator is
\begin{equation*}
\widehat\psi_{\mathrm{RSW}}(z)
=
\frac{
\sum_{i=1}^n
\mathbb I(Z_i=z)\widehat W_i\widehat\tau(z,\widehat W_i)
}{
\sum_{i=1}^n
\mathbb I(Z_i=z)\widehat W_i
}.
\end{equation*}
This empirical analog of \eqref{eq:psi_weighted} learns the nuisance functions and \((Z,W)\)-indexed effect surface in the full sample, then targets the high-risk population by risk-score weighting.

The full-sample-nuisance hard-restriction (FS-HR) benchmark follows \citet{park2024groupwise}: after estimating \(\widehat\Gamma_i\) in the full sample, it averages the signal within the observed high-risk-by-subgroup cells:
\[
\widehat\psi_{\mathrm{FS\text{-}HR}}(z)
=
\frac{
\sum_{i=1}^n
\mathbb I(Z_i=z)H_i\widehat\Gamma_i
}{
\sum_{i=1}^n
\mathbb I(Z_i=z)H_i
}.
\]
Thus, lower-risk observations contribute only to nuisance estimation. The hard-restriction (HR) benchmark instead restricts the entire cross-fitted IRM analysis to \(\mathcal H=\{i:H_i=1\}\), uses \((X_i,Z_i)\) without \(W\), and yields the restricted-sample signal \(\widetilde\Gamma_i\):
\[
\widehat\psi_{\mathrm{HR}}(z)
=
\frac{
\sum_{i\in\mathcal H}
\mathbb I(Z_i=z)\widetilde\Gamma_i
}{
\sum_{i\in\mathcal H}
\mathbb I(Z_i=z)
}.
\]
This is the groupwise estimator of \citet{park2024groupwise} applied to the high-risk subset. Comparing FS-HR with HR isolates full-sample nuisance estimation; comparing RSW with FS-HR isolates soft rather than hard aggregation.

The hard-targeting methods require observed high-risk membership and therefore are available only in Setting I. In Setting II, where $D_i$ and $H_i$ are not observed in the analysis cohort, RSW is the only available estimator. However, if a validation subset with observed $H_i$ is available, FS-HR or HR can be computed as sensitivity analyses to examine the validity of external risk-score assumptions.

\subsection{Inference}
\label{sec:inference}

For Setting~I, let \(\mathrm{IF}_{\psi,\textnormal{I}}(O_i;z)\) denote the mean-zero influence function obtained from the stacked estimating equations for the internally estimated membership score, the treatment-effect projection, and the final RSW aggregation. Let \(\widehat{\mathrm{IF}}_{\psi,\textnormal{I},i}(z)\) be its plug-in estimate.

\begin{theorem}[Inference under Setting I]
\label{thm:rsw_inference_setting1}
For fixed \(z\), suppose Assumptions~(A1)--(A4) and conditions (B1)--(B5) in Appendix~A hold. Then
\begin{equation*}
\sqrt n\{\widehat\psi_{\mathrm{RSW}}(z)-\psi(z)\}
=n^{-1/2}\sum_{i=1}^n\mathrm{IF}_{\psi,\textnormal{I}}(O_i;z)+o_p(1)
\ \rightsquigarrow\ N\{0,V_{\mathrm{RSW},\textnormal{I}}(z)\},
\end{equation*}
where
\begin{equation*}
V_{\mathrm{RSW},\textnormal{I}}(z)=E\{\mathrm{IF}_{\psi,\textnormal{I}}(O_i;z)^2\},\qquad
\widehat V_{\mathrm{RSW},\textnormal{I}}(z)
=n^{-1}\sum_{i=1}^n\widehat{\mathrm{IF}}_{\psi,\textnormal{I},i}(z)^2.
\end{equation*}
The estimator \(\widehat V_{\mathrm{RSW},\textnormal{I}}(z)\) is consistent for \(V_{\mathrm{RSW},\textnormal{I}}(z)\), and hence
\(
\widehat{\mathrm{Var}}\{\widehat\psi_{\mathrm{RSW}}(z)\}
=\widehat V_{\mathrm{RSW},\textnormal{I}}(z)/n.
\)
If \(z_{1-\alpha/2}\) is the \((1-\alpha/2)\)-quantile of the standard normal distribution, a pointwise \((1-\alpha)\) confidence interval is
\(
\widehat\psi_{\mathrm{RSW}}(z)
\ \pm\ z_{1-\alpha/2}\{\widehat V_{\mathrm{RSW},\textnormal{I}}(z)/n\}^{1/2}.
\)
\end{theorem}

For Setting~II, denote the analysis sample and reference sample by \(A\) and \(R\), respectively. For \(s\in\{A,R\}\), write \(W_s(x,z)=P_s(H=1\mid X=x,Z=z)\), where \(P_s\) denotes probability in population \(s\). Let \(\mathrm{IF}_{\psi,\textnormal{II}}^A(O_i;z)\) denote the mean-zero analysis-sample influence function conditional on the population membership score, and let \(\mathrm{IF}_{\psi,\textnormal{II}}^R(O_j;z)\) denote the mean-zero reference-sample influence function induced by estimating that score.

\begin{theorem}[Inference under Setting II]
\label{thm:rsw_inference_setting2}
For each fixed \(z\), suppose Assumptions~(A1)--(A4), conditions (B1), (B2), (B4), and (B5) for the analysis sample, and the reference-sample analog of condition (B3) hold. Further suppose
\begin{equation*}
\begin{gathered}
W_R(x,z)=W_A(x,z)\quad\text{for all }(x,z)\in\operatorname{supp}_A(X,Z),\quad
n,m\to\infty,\quad n/m\longrightarrow\kappa\in(0,\infty).
\end{gathered}
\end{equation*}
Then
\begin{equation*}
\widehat\psi_{\mathrm{RSW}}(z)-\psi(z)
=n^{-1}\sum_{i=1}^{n}\mathrm{IF}_{\psi,\textnormal{II}}^A(O_i;z)
+m^{-1}\sum_{j=1}^{m}\mathrm{IF}_{\psi,\textnormal{II}}^R(O_j;z)
+o_p(n^{-1/2}+m^{-1/2}).
\end{equation*}
Moreover,
\begin{equation*}
\{\widehat\psi_{\mathrm{RSW}}(z)-\psi(z)\}/
\{V_{\mathrm{RSW},\textnormal{II}}^A(z)/n+V_{\mathrm{RSW},\textnormal{II}}^R(z)/m\}^{1/2}
\ \rightsquigarrow\ N(0,1),
\end{equation*}
where \(E_A\) and \(E_R\) denote expectations under the analysis and reference populations, respectively, and
\(
V_{\mathrm{RSW},\textnormal{II}}^A(z)=E_A\{\mathrm{IF}_{\psi,\textnormal{II}}^A(O_i;z)^2\},
V_{\mathrm{RSW},\textnormal{II}}^R(z)=E_R\{\mathrm{IF}_{\psi,\textnormal{II}}^R(O_j;z)^2\}.
\)
The corresponding empirical second moments are
\(
\widehat V_{\mathrm{RSW},\textnormal{II}}^A(z)
=n^{-1}\sum_{i=1}^n\{\widehat{\mathrm{IF}}_{\psi,\textnormal{II},i}^A(z)\}^2,
\widehat V_{\mathrm{RSW},\textnormal{II}}^R(z)
=m^{-1}\sum_{j=1}^m\{\widehat{\mathrm{IF}}_{\psi,\textnormal{II},j}^R(z)\}^2.
\)
A consistent variance estimator is
\(
\widehat{\mathrm{Var}}\{\widehat\psi_{\mathrm{RSW}}(z)\}
=\widehat V_{\mathrm{RSW},\textnormal{II}}^A(z)/n
+\widehat V_{\mathrm{RSW},\textnormal{II}}^R(z)/m.
\)
If \(z_{1-\alpha/2}\) is the \((1-\alpha/2)\)-quantile of the standard normal distribution, a pointwise \((1-\alpha)\) confidence interval is
\(
\widehat\psi_{\mathrm{RSW}}(z)
\ \pm\ z_{1-\alpha/2}
\{\widehat V_{\mathrm{RSW},\textnormal{II}}^A(z)/n
+\widehat V_{\mathrm{RSW},\textnormal{II}}^R(z)/m\}^{1/2}.
\)
\end{theorem}
If the external membership-score model is treated as fixed, the reference-sample influence function is absent and the Setting~II result is obtained by setting \(V_{\mathrm{RSW},\textnormal{II}}^R(z)=0\).

\subsection{Efficiency comparison}
\label{sec:oracle_efficiency}
The following comparison separates two potential sources of precision gain of the proposed estimator. First, suppose that both \(W_i\) and the treatment-effect surface \(\tau(Z_i,W_i)\) are known. To isolate the effect of soft aggregation versus hard restriction, define the oracle hard-restriction and oracle RSW estimators as
\begin{equation*}
\widetilde\psi_H(z)
=\frac{\sum_{i=1}^n\mathbb I(Z_i=z)H_i\tau(z,W_i)}
{\sum_{i=1}^n\mathbb I(Z_i=z)H_i},\qquad
\widetilde\psi_W(z)
=\frac{\sum_{i=1}^n\mathbb I(Z_i=z)W_i\tau(z,W_i)}
{\sum_{i=1}^n\mathbb I(Z_i=z)W_i}.
\end{equation*}
Under \(E(H_i\mid W_i,Z_i)=W_i\), both target to estimate \(\psi(z)\). Write \(G_{i,z}=\mathbb I(Z_i=z)\), \(\Delta_i(z)=\tau(z,W_i)-\psi(z)\), and
$
q_z=E(G_{i,z}H_i)=E(G_{i,z}W_i).
$
The corresponding influence functions are
\begin{equation*}
\phi_{H,i}(z)
=\frac{G_{i,z}H_i\Delta_i(z)}{q_z},\qquad
\phi_{W,i}(z)
=\frac{G_{i,z}W_i\Delta_i(z)}{q_z}.
\end{equation*}
Consequently,
\begin{equation*}
\mathrm{Var}\{\phi_{H,i}(z)\}-\mathrm{Var}\{\phi_{W,i}(z)\}
=\frac{E\{G_{i,z}W_i(1-W_i)\Delta_i(z)^2\}}{q_z^2}\geq0.
\end{equation*}
This result suggests that replacing the hard membership indicator by its conditional probability can reduce asymptotic
variance, which is a Rao-Blackwell-type smoothing effect. The inequality is strict whenever there are subjects with $0<W_i<1$ and treatment effects vary with $W_i$.

However, \(\widetilde\psi_H(z)\) and \(\widetilde\psi_W(z)\) are not the complete FS-HR and RSW estimators; they are auxiliary oracle estimators that isolate their respective hard- and soft-aggregation steps. In particular, they omit the residual variation in the orthogonal signal and, for RSW, the variation from estimating the treatment-effect projection. Additionally, suppose that \(\tau(Z_i,W_i)=b(Z_i,W_i)^\top\beta_0\) is correctly specified and, with \(\varepsilon_i=\Gamma_i(\eta_0)-\tau(Z_i,W_i)\),
\begin{equation*}
E(\varepsilon_i\mid H_i,W_i,Z_i)=0,\qquad
\mathrm{Var}(\varepsilon_i\mid H_i,W_i,Z_i)=\sigma_\varepsilon^2.
\end{equation*}
Define \(b_i=b(Z_i,W_i)\), \(Q=E(b_ib_i^\top)\), and \(c_z=E(G_{i,z}W_i b_i)\), and suppose that \(Q\) is nonsingular. Let \(\widehat\psi_{\mathrm{RSW}}^{*}(z)\) denote the complete RSW estimator with \(W\) treated as known but \(\beta_0\) still estimated from the orthogonal signal. The influence functions of FS-HR and this known-score RSW estimator are
\begin{align*}
\mathrm{IF}_{\mathrm{FS\text{-}HR},i}(z)
=q_z^{-1}G_{i,z}H_i\{\Delta_i(z)+\varepsilon_i\},\quad
\mathrm{IF}_{\mathrm{RSW},i}^{*}(z)
=q_z^{-1}\{G_{i,z}W_i\Delta_i(z)+c_z^\top Q^{-1}b_i\varepsilon_i\}.
\end{align*}
Let \(V_{\mathrm{FS\text{-}HR}}(z)\) and \(V_{\mathrm{RSW}}^{*}(z)\) denote their respective asymptotic variances. Then
\begin{align*}
V_{\mathrm{FS\text{-}HR}}(z)-V_{\mathrm{RSW}}^{*}(z)
=\frac{1}{q_z^2}\big[&E\{G_{i,z}W_i(1-W_i)\Delta_i(z)^2\}\\
&+\sigma_\varepsilon^2\{q_z-c_z^\top Q^{-1}c_z\}\big]\geq0.
\end{align*}
The first term is the soft-versus-hard aggregation gain identified by \(\widetilde\psi_H(z)\) and \(\widetilde\psi_W(z)\). The second is the projection gain from estimating a correctly specified, low-dimensional treatment-effect surface with the full sample. Because \(W\) is treated as known, this comparison is common to Settings~I and~II.

When $W$ is estimated instead of known from prior knowledge, the variance comparison is less direct. In Setting~I, estimating $W$ from the analysis sample introduces an additional first-order influence-function contribution that may be correlated with $\mathrm{IF}_{\mathrm{RSW},i}^{*}(z)$. Thus, the variance reduction from soft targeting may be partially offset by uncertainty from estimating the membership score, and uniform efficiency gain over FS-HR is not generally guaranteed. In Setting~II, estimating $W$ in an independent reference sample similarly contributes the additional variance term $V_{\mathrm{RSW},\textnormal{II}}^R(z)/m$ in Theorem~\ref{thm:rsw_inference_setting2}. Thus, a simple general efficiency gain need not hold. Moreover, because $H$ is unobserved in the Setting~II analysis sample, FS-HR is not directly available unless high-risk status is observed in a separate validation sample.

\section{Simulation studies}

\subsection{Main simulation design}
We generated a four-level subgroup variable $Z\in\{1,\ldots,4\}$ with $\Pr(Z=1)=0.40$, $\Pr(Z=2)=0.20$, $\Pr(Z=3)=0.25$, $\Pr(Z=4)=0.15.$ Baseline covariates $X=(X_1,\ldots,X_7)$ were generated independently as $X_1\sim N(59.2,2.7^2)$, $X_2\sim N(27.5,3.3^2)$, $X_3\sim N(50.7,16.7^2)$, $X_4\sim N(211.8,41.1^2)$, $X_5\sim \mathrm{Gamma}(6,0.4)$, $X_6\sim \mathrm{Bernoulli}(0.546)$, and $X_7\in\{1,2,3\}$ with probabilities $(0.233,0.411,0.356)$.

The observed ordinal risk measure \(D\in\{0,1,2\}\) was generated from a latent logistic model with \(\eta_D=0.5X_1+0.4X_2+0.5X_3+0.3X_4+0.6X_5+0.1Z-140\) and \(S_D=\eta_D+\varepsilon_D\), where \(\varepsilon_D\sim\mathrm{Logistic}(0,1)\). Setting \(c_1=\log(3/14)\), we defined \(D=0\) if \(S_D\le c_1\), \(D=1\) if \(c_1<S_D\le0\), and \(D=2\) if \(S_D>0\). Thus, \(W(X,Z)=\Pr(D>1\mid X,Z)=\operatorname{expit}\{\eta_D\}\).

Treatment was assigned by $A\sim \mathrm{Bernoulli}\{\pi_A\}$, where
$\pi_A=\operatorname{expit}\{(\eta_A-3)/3\}$, $\eta_A=0.025X_1+0.05X_2+0.1X_3-0.025X_4+0.15X_5
+0.10X_6+0.10 I(X_7=3)+0.02(Z-2)^3$.

We considered continuous and binary outcomes, and for each outcome type we examined both linear and nonlinear treatment-effect surfaces in the membership score $W$.

\paragraph{(1) Continuous outcome} For continuous outcomes, we set
\(
\mu_c=0.6X_1+0.4X_3-0.2X_4+X_5I(X_6=1)+0.6 I(X_7=3)+\log (Z-0.3).
\)
The linear and nonlinear treatment effects were $\tau_c^{\mathrm{lin}}=Z+8W-5$ and $\tau_c^{\mathrm{nonlin}}=2\sqrt{Z}-4W(W-1)(W+6)$, respectively. We generated $Y\sim\mathrm{Normal}\left(\mu_c+A\tau_c,1\right)$.

\paragraph{(2) Binary outcome} For binary outcomes, we worked on the risk scale, and the baseline risk was \(\mu_b=0.12+0.003X_2+0.05Z\). The linear and nonlinear treatment effects were $\tau_b^{\mathrm{lin}}=0.1Z+0.1W$ and $\tau_b^{\mathrm{nonlin}}=0.03Z^2+0.06W(W-1)(W-3)$. We generated $Y\sim\mathrm{Bernoulli}\{\mu_b+A\tau_b\}$.

We considered two sample sizes, $n=500$ and $n=2000$. Within each sample size, we crossed two outcome types, continuous and binary, with two treatment-effect settings, linear and nonlinear in the membership score $W$. This yielded eight main simulation scenarios.

We evaluated both risk-status settings under this common data-generating mechanism. In Setting~I, each replicate consisted of one analysis sample with observed \(H\), and the membership-score model was estimated within the analysis-sample cross-fitting folds. In Setting~II, each replicate also included an independent reference sample generated from the same mechanism. We considered reference-sample sizes \(m=n/4\) and \(m=2n\); in each case, the membership-score model was fitted in the reference sample, and its frozen predictions were applied to the analysis sample. The analysis-sample \(H\) was retained to compute FS-HR and HR as oracle benchmarks.

The Monte Carlo truth was defined using an independent sample of size \(10^6\) generated from the same data-generating mechanism. We evaluated the individual conditional treatment effect \(\tau_i=E\{Y_i(1)-Y_i(0)\mid X_i,Z_i\}\) from the known outcome model and computed the truth as the average of \(\tau_i\) among individuals with \(H_i=1\) and \(Z_i=z\).

All three estimators used the IRM framework. The treatment nuisance function was estimated by \(L_2\)-penalized logistic regression, and the outcome nuisance function was estimated by ridge regression for continuous outcomes and \(L_2\)-penalized logistic regression for binary outcomes. We used two-fold cross-fitting, with folds stratified by treatment status. The RSW basis matched the treatment-effect setting: subgroup indicators and a linear term in \(W\) for the linear setting, and subgroup indicators plus a common B-spline basis in \(W\) with 6 degrees of freedom and degree 4 for the nonlinear setting.

Across $R=500$ replications in each scenario, we summarized each subgroup-specific estimator by its Monte Carlo mean, absolute bias, empirical standard deviation (SD), mean squared error (MSE), and empirical coverage of nominal 95\% confidence intervals.

\subsection{Main simulation results}

\begin{figure}[H]
\centering
\includegraphics[width=\textwidth,alt={Across twelve continuous-outcome boxplot panels, the risk-score-weighted estimator generally has the narrowest distributions and the hard-restriction estimator the widest, while all estimators remain centered near the truths.}]{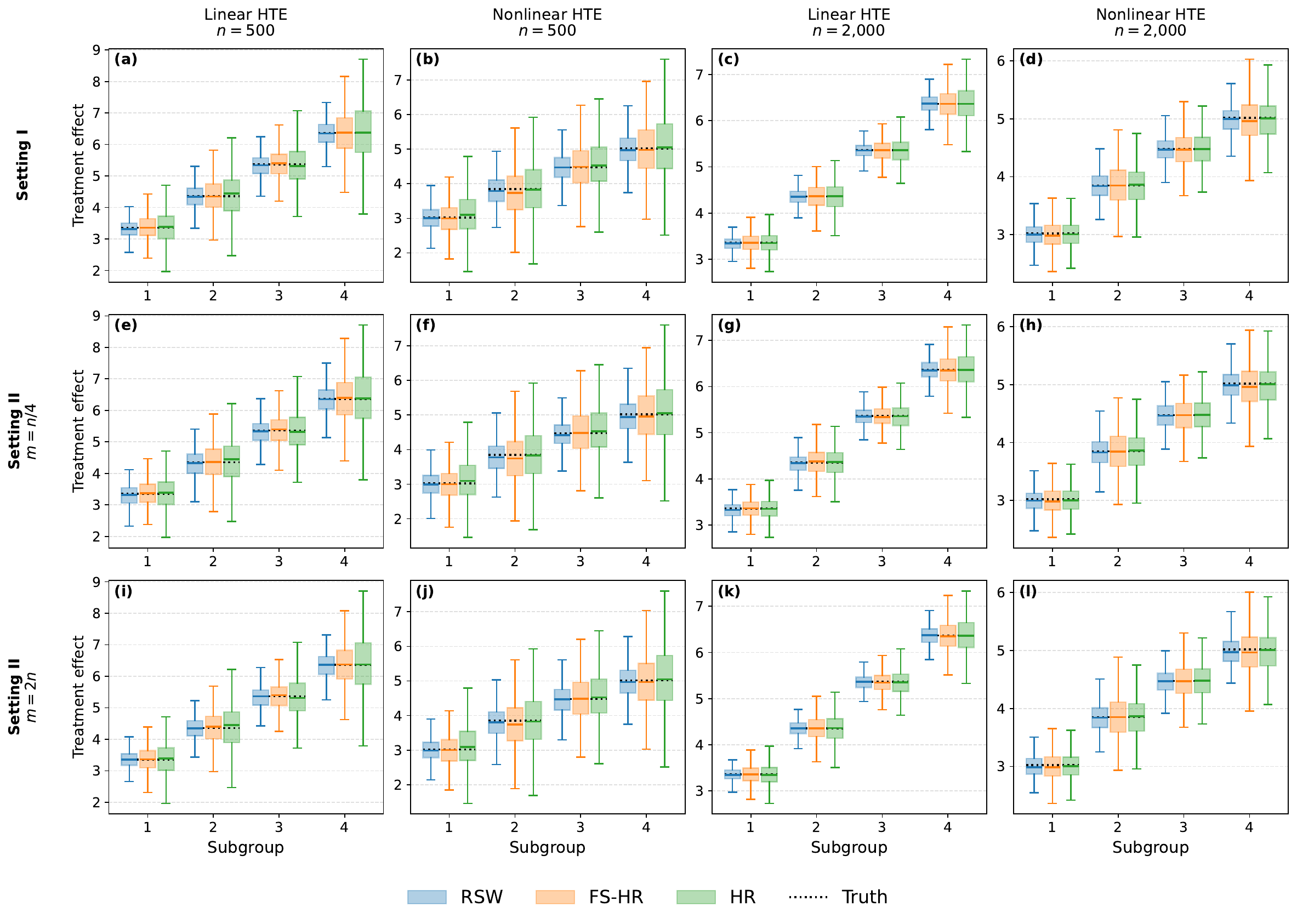}

\caption{Main simulation results for continuous outcomes under Settings~I and~II. Boxplots show the Monte Carlo distributions of RSW, FS-HR, and HR across 500 replications. Rows show Setting~I, Setting~II with \(m=n/4\), and Setting~II with \(m=2n\); columns show linear and nonlinear HTE at \(n=500\) and \(n=2000\). Dotted lines denote the Monte Carlo truth \(\psi_0(z)\).}
\label{fig:main-sim-continuous-boxplots}
\end{figure}

\begin{figure}[H]
\centering
\includegraphics[width=\textwidth,alt={Across twelve binary-outcome boxplot panels, the risk-score-weighted estimator generally has the narrowest distributions and the hard-restriction estimator the widest, while all estimators remain centered near the truths.}]{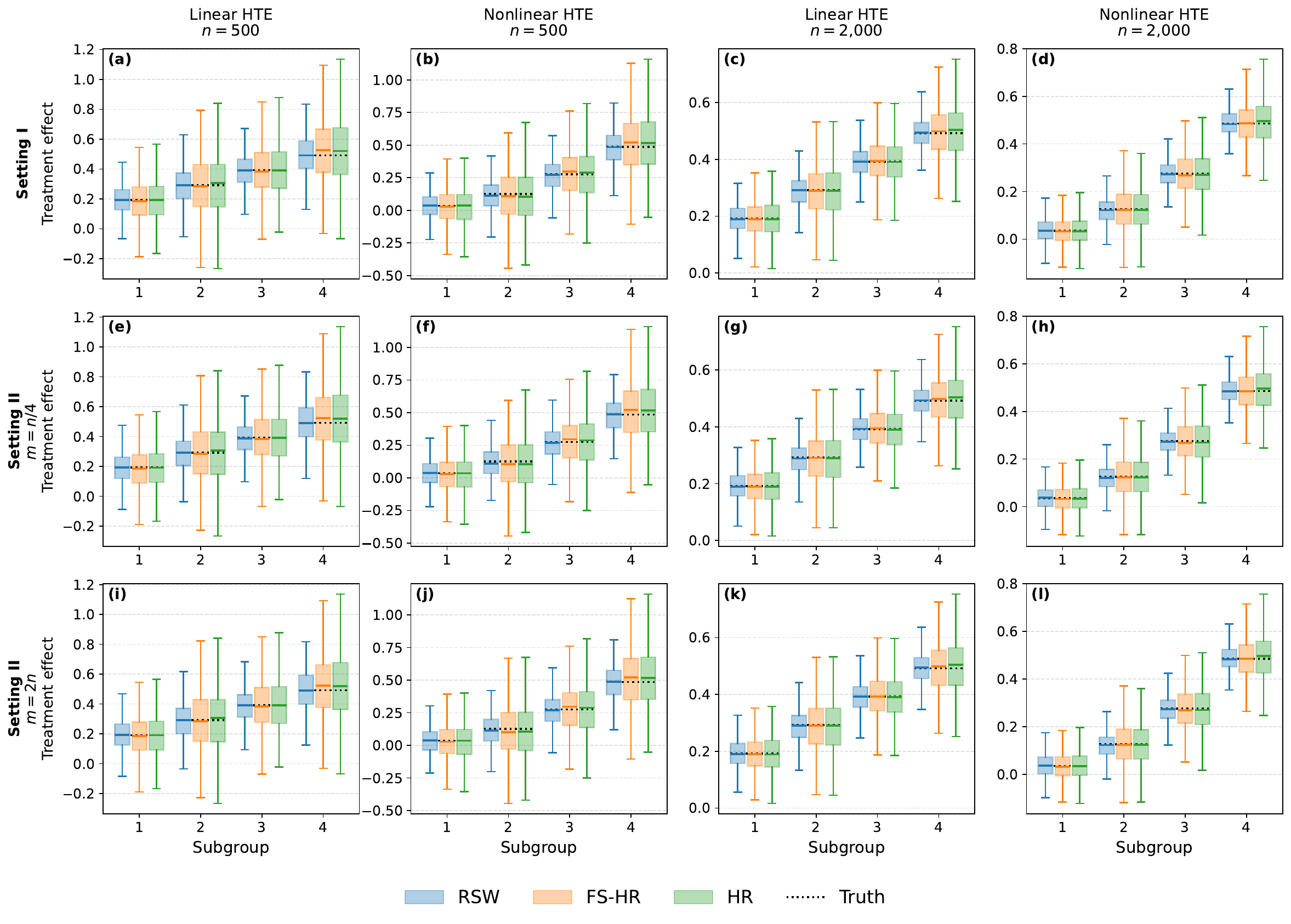}

\caption{Main simulation results for binary outcomes on the risk-difference scale under Settings~I and~II. Boxplots show the Monte Carlo distributions of RSW, FS-HR, and HR across 500 replications. Rows show Setting~I, Setting~II with \(m=n/4\), and Setting~II with \(m=2n\); columns show linear and nonlinear HTE at \(n=500\) and \(n=2000\). Dotted lines denote the Monte Carlo truth \(\psi_0(z)\).}
\label{fig:main-sim-binary-boxplots}
\end{figure}

Across both settings, reference-sample sizes, and all eight outcome, sample-size, and HTE scenarios, the three estimators were generally centered near the Monte Carlo truth and empirical coverage was close to the nominal 95\% level. RSW had the smallest empirical SD throughout, with FS-HR generally intermediate and HR the most variable; these differences were largest at \(n=500\). Figures~\ref{fig:main-sim-continuous-boxplots} and~\ref{fig:main-sim-binary-boxplots} show the continuous- and binary-outcome distributions, respectively. For example, for continuous outcomes with linear HTE at \(n=500\), the subgroup-averaged RSW SD decreased from 0.416 with \(m=n/4\) to 0.348 with \(m=2n\), compared with 0.569 for FS-HR and 0.901 for HR under \(m=n/4\). The effect of increasing the reference-sample size was smaller under nonlinear HTE and negligible for binary outcomes. At \(n=2000\), all estimators became more stable, and the difference between the hard-targeting estimators narrowed. Full subgroup-specific summaries are provided in Appendix~B and Supplementary Tables~1--6.

\subsection{Sensitivity analysis}
The main simulation evaluates performance under the treatment-effect sufficiency assumption (A4). We conducted a sensitivity analysis to assess how violations of this condition affect the performance of estimators.
We retained the same distributions for $(X,Z,A)$ as in the main analysis. We additionally introduced an unobserved independent binary variable $U$ with $\Pr(U=\pm1)=1/2$.
Let \(\rho\in\{0,0.3,0.6,0.9\}\) be the sensitivity parameter. For a given \(\rho\), high-risk membership was generated so that
\[
\Pr(D>1\mid X,Z,U)=\operatorname{expit}\{\eta_D\}+\rho U \operatorname{expit}\{\eta_D\}\{1-\operatorname{expit}\{\eta_D\}\}.
\]
Marginalizing over \(U\) leaves the observed-variable score unchanged: \(W(X,Z)=\operatorname{expit}\{\eta_D\}\).

\paragraph{(1) Continuous outcome} We modified the baseline outcome and treatment effect as
\[
\mu_c^\rho=\mu_c+\rho U,
\qquad
\tau_c^\rho=\tau_c+9\rho U,
\]
where \(\mu_c\) and \(\tau_c\) are from the main simulation. We generated $Y\sim N\{\mu_c^\rho+A\tau_c^\rho,1\}$.

\paragraph{(2) Binary outcome} We modified the baseline risk and risk-difference treatment
effect as
\[
\mu_b^\rho=\mu_b-0.01\rho U,
\qquad
\tau_b^\rho=\tau_b+0.05\rho U,
\]
where \(\mu_b\) and \(\tau_b\) are from the main simulation. We generated $Y\sim \mathrm{Bernoulli}\{\mu_b^\rho+A\tau_b^\rho\}$.

When \(\rho=0\), these models reduce to the outcome models in the main simulation. When \(\rho>0\), high-risk membership \(H\) is informative about \(U\) even after conditioning on \((W,Z)\). Because \(U\) also modifies the potential outcomes, \(Y(a)\not\perp\!\!\!\perp H\mid W,Z\). Thus, \(\rho>0\) creates a controlled violation of Assumption~(A4) while leaving the observed-variable membership score unchanged.

For each value of \(\rho\), we repeated the same simulation branches: Setting~I estimated \(W\) internally from the analysis sample, whereas Setting~II estimated \(W\) in an independent reference sample of size \(m=n/4\) or \(m=2n\) generated under the corresponding sensitivity mechanism.

For each value of \(\rho\), the Monte Carlo truth used the same direct high-risk estimand as in the main simulation: in an independent sample of size \(10^6\), we averaged the known treatment effect \(\tau_i^\rho\) among individuals with \(H_i=1\) within each subgroup \(z\). The analysis used \(n=2000\) and the same continuous and binary outcome types and linear and nonlinear treatment-effect surfaces as in the main simulation. The sensitivity analyses also used two-fold cross-fitting.

\begin{table}[p]
\centering
\caption{Sensitivity analysis for continuous outcomes under violations of treatment-effect sufficiency, with analysis-sample size $n=2000$. The sensitivity parameter $\rho$ controls the strength of residual dependence between high-risk membership and the unobserved effect modifier $U$ after conditioning on $(W,Z)$. Absolute bias, empirical Monte Carlo SD, MSE, and coverage of nominal 95\% Wald confidence intervals are averaged over the four subgroup-specific targets across 500 replications. The three vertically stacked blocks report Setting~I, Setting~II with $m=n/4$, and Setting~II with $m=2n$.}
\label{tab:sens-continuous}
\begingroup
\footnotesize
\setlength{\tabcolsep}{1.5pt}
\renewcommand{\arraystretch}{0.82}
\begin{tabular*}{\textwidth}{@{\extracolsep{\fill}}l r r r r r r r r r r r r @{}}
\toprule
& \multicolumn{4}{c}{$\widehat{\psi}_{\mathrm{RSW}}$} & \multicolumn{4}{c}{$\widehat{\psi}_{\mathrm{FS\text{-}HR}}$} & \multicolumn{4}{c}{$\widehat{\psi}_{\mathrm{HR}}$} \\
\cmidrule(lr){2-5}\cmidrule(lr){6-9}\cmidrule(lr){10-13}
$\rho$ & Abs. bias & SD & MSE & Cov. & Abs. bias & SD & MSE & Cov. & Abs. bias & SD & MSE & Cov. \\
\midrule
\multicolumn{13}{l}{\textit{Setting I}} \\
\addlinespace[0.10em]
\multicolumn{13}{l}{\textit{Panel A. Linear HTE}} \\
\addlinespace[0.08em]
0 & 0.010 & 0.179 & 0.032 & 0.940 & 0.008 & 0.257 & 0.068 & 0.934 & 0.011 & 0.296 & 0.090 & 0.934 \\
0.3 & 0.066 & 0.318 & 0.107 & 0.942 & 0.026 & 0.477 & 0.235 & 0.947 & 0.019 & 0.508 & 0.266 & 0.938 \\
0.6 & 0.243 & 0.556 & 0.373 & 0.920 & 0.017 & 0.841 & 0.725 & 0.944 & 0.024 & 0.873 & 0.783 & 0.940 \\
0.9 & 0.600 & 0.821 & 1.043 & 0.876 & 0.027 & 1.228 & 1.560 & 0.939 & 0.038 & 1.255 & 1.626 & 0.936 \\
\addlinespace[0.12em]
\multicolumn{13}{l}{\textit{Panel B. Nonlinear HTE}} \\
\addlinespace[0.08em]
0 & 0.020 & 0.218 & 0.049 & 0.956 & 0.031 & 0.323 & 0.110 & 0.952 & 0.009 & 0.313 & 0.102 & 0.948 \\
0.3 & 0.067 & 0.345 & 0.126 & 0.942 & 0.025 & 0.531 & 0.291 & 0.946 & 0.023 & 0.532 & 0.293 & 0.944 \\
0.6 & 0.240 & 0.590 & 0.410 & 0.916 & 0.021 & 0.898 & 0.825 & 0.944 & 0.047 & 0.912 & 0.851 & 0.938 \\
0.9 & 0.610 & 0.837 & 1.083 & 0.876 & 0.032 & 1.290 & 1.726 & 0.937 & 0.025 & 1.307 & 1.764 & 0.938 \\
\midrule
\multicolumn{13}{l}{\textit{Setting II: $m=n/4$}} \\
\addlinespace[0.10em]
\multicolumn{13}{l}{\textit{Panel A. Linear HTE}} \\
\addlinespace[0.08em]
0 & 0.013 & 0.202 & 0.042 & 0.947 & 0.008 & 0.259 & 0.069 & 0.938 & 0.011 & 0.296 & 0.090 & 0.934 \\
0.3 & 0.068 & 0.327 & 0.113 & 0.952 & 0.028 & 0.477 & 0.235 & 0.949 & 0.019 & 0.508 & 0.266 & 0.938 \\
0.6 & 0.244 & 0.565 & 0.383 & 0.930 & 0.018 & 0.841 & 0.726 & 0.942 & 0.024 & 0.873 & 0.783 & 0.940 \\
0.9 & 0.608 & 0.828 & 1.067 & 0.880 & 0.028 & 1.228 & 1.558 & 0.939 & 0.038 & 1.255 & 1.626 & 0.936 \\
\addlinespace[0.12em]
\multicolumn{13}{l}{\textit{Panel B. Nonlinear HTE}} \\
\addlinespace[0.08em]
0 & 0.023 & 0.229 & 0.054 & 0.950 & 0.031 & 0.322 & 0.110 & 0.954 & 0.009 & 0.313 & 0.102 & 0.948 \\
0.3 & 0.070 & 0.349 & 0.129 & 0.946 & 0.026 & 0.531 & 0.291 & 0.947 & 0.023 & 0.532 & 0.293 & 0.944 \\
0.6 & 0.245 & 0.594 & 0.416 & 0.920 & 0.021 & 0.897 & 0.824 & 0.945 & 0.047 & 0.912 & 0.851 & 0.938 \\
0.9 & 0.614 & 0.838 & 1.090 & 0.881 & 0.031 & 1.291 & 1.727 & 0.939 & 0.025 & 1.307 & 1.764 & 0.938 \\
\midrule
\multicolumn{13}{l}{\textit{Setting II: $m=2n$}} \\
\addlinespace[0.10em]
\multicolumn{13}{l}{\textit{Panel A. Linear HTE}} \\
\addlinespace[0.08em]
0 & 0.010 & 0.171 & 0.030 & 0.946 & 0.007 & 0.256 & 0.067 & 0.936 & 0.011 & 0.296 & 0.090 & 0.934 \\
0.3 & 0.068 & 0.311 & 0.102 & 0.952 & 0.027 & 0.476 & 0.234 & 0.945 & 0.019 & 0.508 & 0.266 & 0.938 \\
0.6 & 0.241 & 0.555 & 0.370 & 0.924 & 0.017 & 0.839 & 0.723 & 0.944 & 0.024 & 0.873 & 0.783 & 0.940 \\
0.9 & 0.598 & 0.819 & 1.040 & 0.884 & 0.027 & 1.229 & 1.563 & 0.938 & 0.038 & 1.255 & 1.626 & 0.936 \\
\addlinespace[0.12em]
\multicolumn{13}{l}{\textit{Panel B. Nonlinear HTE}} \\
\addlinespace[0.08em]
0 & 0.018 & 0.217 & 0.048 & 0.948 & 0.031 & 0.323 & 0.110 & 0.952 & 0.009 & 0.313 & 0.102 & 0.948 \\
0.3 & 0.066 & 0.344 & 0.125 & 0.946 & 0.025 & 0.531 & 0.292 & 0.946 & 0.023 & 0.532 & 0.293 & 0.944 \\
0.6 & 0.238 & 0.589 & 0.408 & 0.922 & 0.021 & 0.898 & 0.824 & 0.945 & 0.047 & 0.912 & 0.851 & 0.938 \\
0.9 & 0.606 & 0.835 & 1.075 & 0.878 & 0.031 & 1.291 & 1.727 & 0.938 & 0.025 & 1.307 & 1.764 & 0.938 \\
\bottomrule
\end{tabular*}
\endgroup
\end{table}

\begin{table}[p]
\centering
\caption{Sensitivity analysis for binary outcomes under violations of treatment-effect sufficiency, with analysis-sample size $n=2000$. The sensitivity parameter $\rho$ controls the strength of residual dependence between high-risk membership and the unobserved effect modifier $U$ after conditioning on $(W,Z)$. Absolute bias, empirical Monte Carlo SD, MSE, and coverage of nominal 95\% Wald confidence intervals are averaged over the four subgroup-specific targets across 500 replications. The three vertically stacked blocks report Setting~I, Setting~II with $m=n/4$, and Setting~II with $m=2n$.}
\label{tab:sens-binary}
\begingroup
\footnotesize
\setlength{\tabcolsep}{1.5pt}
\renewcommand{\arraystretch}{0.82}
\begin{tabular*}{\textwidth}{@{\extracolsep{\fill}}l r r r r r r r r r r r r @{}}
\toprule
& \multicolumn{4}{c}{$\widehat{\psi}_{\mathrm{RSW}}$} & \multicolumn{4}{c}{$\widehat{\psi}_{\mathrm{FS\text{-}HR}}$} & \multicolumn{4}{c}{$\widehat{\psi}_{\mathrm{HR}}$} \\
\cmidrule(lr){2-5}\cmidrule(lr){6-9}\cmidrule(lr){10-13}
$\rho$ & Abs. bias & SD & MSE & Cov. & Abs. bias & SD & MSE & Cov. & Abs. bias & SD & MSE & Cov. \\
\midrule
\multicolumn{13}{l}{\textit{Setting I}} \\
\addlinespace[0.10em]
\multicolumn{13}{l}{\textit{Panel A. Linear HTE}} \\
\addlinespace[0.08em]
0 & 0.002 & 0.059 & 0.004 & 0.936 & 0.004 & 0.087 & 0.008 & 0.941 & 0.006 & 0.090 & 0.008 & 0.946 \\
0.3 & 0.004 & 0.055 & 0.003 & 0.959 & 0.005 & 0.084 & 0.007 & 0.951 & 0.005 & 0.088 & 0.008 & 0.956 \\
0.6 & 0.001 & 0.056 & 0.003 & 0.956 & 0.001 & 0.088 & 0.008 & 0.939 & 0.001 & 0.092 & 0.009 & 0.942 \\
0.9 & 0.002 & 0.057 & 0.003 & 0.951 & 0.002 & 0.086 & 0.008 & 0.947 & 0.003 & 0.090 & 0.008 & 0.944 \\
\addlinespace[0.12em]
\multicolumn{13}{l}{\textit{Panel B. Nonlinear HTE}} \\
\addlinespace[0.08em]
0 & 0.002 & 0.058 & 0.003 & 0.947 & 0.003 & 0.088 & 0.008 & 0.940 & 0.005 & 0.092 & 0.009 & 0.938 \\
0.3 & 0.003 & 0.056 & 0.003 & 0.962 & 0.005 & 0.085 & 0.007 & 0.950 & 0.005 & 0.090 & 0.008 & 0.949 \\
0.6 & 0.001 & 0.056 & 0.003 & 0.956 & 0.003 & 0.086 & 0.008 & 0.946 & 0.003 & 0.090 & 0.008 & 0.951 \\
0.9 & 0.004 & 0.058 & 0.003 & 0.942 & 0.003 & 0.086 & 0.008 & 0.944 & 0.003 & 0.091 & 0.009 & 0.943 \\
\midrule
\multicolumn{13}{l}{\textit{Setting II: $m=n/4$}} \\
\addlinespace[0.10em]
\multicolumn{13}{l}{\textit{Panel A. Linear HTE}} \\
\addlinespace[0.08em]
0 & 0.002 & 0.059 & 0.004 & 0.942 & 0.004 & 0.087 & 0.008 & 0.942 & 0.006 & 0.090 & 0.008 & 0.946 \\
0.3 & 0.004 & 0.055 & 0.003 & 0.961 & 0.005 & 0.084 & 0.007 & 0.951 & 0.005 & 0.088 & 0.008 & 0.956 \\
0.6 & 0.001 & 0.056 & 0.003 & 0.956 & 0.001 & 0.088 & 0.008 & 0.939 & 0.001 & 0.092 & 0.009 & 0.942 \\
0.9 & 0.002 & 0.058 & 0.003 & 0.949 & 0.003 & 0.086 & 0.007 & 0.948 & 0.003 & 0.090 & 0.008 & 0.944 \\
\addlinespace[0.12em]
\multicolumn{13}{l}{\textit{Panel B. Nonlinear HTE}} \\
\addlinespace[0.08em]
0 & 0.002 & 0.058 & 0.003 & 0.944 & 0.003 & 0.088 & 0.008 & 0.940 & 0.005 & 0.092 & 0.009 & 0.938 \\
0.3 & 0.003 & 0.056 & 0.003 & 0.960 & 0.005 & 0.085 & 0.007 & 0.949 & 0.005 & 0.090 & 0.008 & 0.949 \\
0.6 & 0.001 & 0.056 & 0.003 & 0.962 & 0.003 & 0.086 & 0.008 & 0.946 & 0.003 & 0.090 & 0.008 & 0.951 \\
0.9 & 0.004 & 0.058 & 0.003 & 0.942 & 0.003 & 0.086 & 0.008 & 0.942 & 0.003 & 0.091 & 0.009 & 0.943 \\
\midrule
\multicolumn{13}{l}{\textit{Setting II: $m=2n$}} \\
\addlinespace[0.10em]
\multicolumn{13}{l}{\textit{Panel A. Linear HTE}} \\
\addlinespace[0.08em]
0 & 0.002 & 0.059 & 0.004 & 0.938 & 0.004 & 0.087 & 0.008 & 0.942 & 0.006 & 0.090 & 0.008 & 0.946 \\
0.3 & 0.004 & 0.056 & 0.003 & 0.959 & 0.005 & 0.084 & 0.007 & 0.952 & 0.005 & 0.088 & 0.008 & 0.956 \\
0.6 & 0.001 & 0.056 & 0.003 & 0.956 & 0.001 & 0.088 & 0.008 & 0.939 & 0.001 & 0.092 & 0.009 & 0.942 \\
0.9 & 0.002 & 0.057 & 0.003 & 0.950 & 0.003 & 0.086 & 0.008 & 0.947 & 0.003 & 0.090 & 0.008 & 0.944 \\
\addlinespace[0.12em]
\multicolumn{13}{l}{\textit{Panel B. Nonlinear HTE}} \\
\addlinespace[0.08em]
0 & 0.002 & 0.058 & 0.003 & 0.945 & 0.003 & 0.088 & 0.008 & 0.940 & 0.005 & 0.092 & 0.009 & 0.938 \\
0.3 & 0.003 & 0.056 & 0.003 & 0.958 & 0.005 & 0.085 & 0.007 & 0.950 & 0.005 & 0.090 & 0.008 & 0.949 \\
0.6 & 0.001 & 0.055 & 0.003 & 0.959 & 0.003 & 0.086 & 0.008 & 0.946 & 0.003 & 0.090 & 0.008 & 0.951 \\
0.9 & 0.004 & 0.057 & 0.003 & 0.941 & 0.003 & 0.086 & 0.008 & 0.942 & 0.003 & 0.091 & 0.009 & 0.943 \\
\bottomrule
\end{tabular*}
\endgroup
\end{table}

For continuous outcomes, increasing \(\rho\) increased RSW bias and reduced coverage in both settings, whereas FS-HR and HR retained smaller bias and near-nominal coverage but had greater variability. At \(\rho=0.9\), RSW coverage was approximately 0.88 under both linear and nonlinear HTE. Nevertheless, RSW remained the most efficient estimator and had the lowest MSE throughout the sensitivity scenarios examined. For binary outcomes, all three estimators retained small bias and near-nominal coverage.

Setting~II closely paralleled Setting~I at both reference-sample sizes. A larger reference sample improved RSW precision primarily when \(\rho\) was small; as the violation strengthened, bias dominated and increasing \(m\) could not remedy the violation of treatment-effect sufficiency.

\clearpage
\section{Real-world Analysis in the All\textit{of}Us Cohort}

We applied the proposed framework to the All\textit{of}Us Research Program, a large U.S. cohort linking electronic health records with participant-reported lifestyle and social determinants of health (SDOH) measures \citep{AllofUs2019}. We selected incident mild cognitive impairment (MCI) as a clinically identifiable early cognitive endpoint relevant to progression along the AD continuum \citep{Albert2011MCI}. Motivated by evidence that blood-pressure control and glucose-lowering therapies may influence cognitive outcomes \citep{CukiermanYaffe2020REWIND,Williamson2019}, we estimated the effects of baseline non-GLP antidiabetic and antihypertensive medication use on time to incident MCI.

For each medication-specific analysis, \(A_i\) denoted binary use of the target medication. The baseline adjustment set \(X_i\) comprised age, gender, race, ethnicity, education, seven clinical comorbidities, lifestyle and housing measures, 11 average SDOH scores, and use of nontarget medications. The subgroup variable \(Z_i\) represented four multimodal clusters previously derived from comorbidities, lifestyle factors, and SDOH \citep{Zhao2025.05.27.25328416}; their profiles are summarized in Table~\ref{tab:cluster_profiles}.

\begin{table}[htbp]
\centering
\caption{Description of the four multimodal clusters used as subgroup definitions in the All\textit{of}Us analysis.}
\label{tab:cluster_profiles}
\small
\begin{tabular}{p{0.10\linewidth} p{0.84\linewidth}}
\toprule
\textbf{Cluster} & \textbf{Profile description} \\
\midrule
I & Lowest overall comorbidity burden and comparatively favorable social conditions, consistent with a low-risk healthy-aging profile. \\
II & Greater behavioral and psychosocial adversity, including heavier smoking/alcohol burden and less favorable psychosocial conditions, with less concentrated cardiometabolic disease than Clusters~III and~IV. \\
III & Concentrated cardiometabolic and depressive multimorbidity, with elevated burdens of hypertension, type 2 diabetes, depression, hearing loss, and related chronic disease. \\
IV & Greater social and material vulnerability, including food insecurity, housing problems, and neighborhood adversity, accompanied by elevated cardiometabolic burden. \\
\bottomrule
\end{tabular}
\end{table}

Missing baseline covariates were imputed using multiple imputation by chained equations (MICE), yielding five completed datasets.
All subsequent analyses were conducted separately in each completed dataset and combined using Rubin's rules \citep{Rubin1987}.

Let \(T_i\) denote the time from the survey start date to incident MCI and \(C_i\) the time from the survey start date to the end of observed follow-up. The observed time was \(U_i=\min(T_i,C_i)\), and \(\Delta_i=1\{T_i\le C_i\}\) indicated an observed MCI event. We clinically specified \(t_0=48\) months as a medium-term horizon that allowed time for incident MCI events to accrue while retaining adequate follow-up support and avoiding extrapolation into a sparsely observed longer-term tail. The fixed-horizon outcome was \(Y_i=\log\{\min(T_i,t_0)\}\).

The restricted outcome was observed for participants who developed MCI by \(t_0\) or remained under observation through \(t_0\). We therefore defined \(R_i=1\{(\Delta_i=1\text{ and }U_i\le t_0)\text{ or }U_i\ge t_0\}\). For \(R_i=1\), the outcome was calculated as \(Y_i=\log\{\min(U_i,t_0)\}\); participants with \(R_i=0\) ended follow-up before \(t_0\) without recorded MCI and therefore had an unknown restricted outcome.

To account for selection induced by censoring before the restricted outcome could be observed, we used inverse probability of censoring weights (IPCW) \citep{RobinsFinkelstein2000}. Let \(G(s\mid A_i,X_i,Z_i)=\Pr(C_i\ge s\mid A_i,X_i,Z_i)\) denote the conditional probability of remaining under observation through time \(s\). For the censoring model, we defined \(\widetilde U_i=\min(U_i,t_0)\) and \(\widetilde{\Delta}_i=1\{\Delta_i=0,U_i<t_0\}\), where \(\widetilde{\Delta}_i=1\) indicated that follow-up ended before \(t_0\) without a preceding MCI diagnosis.

Participants with \(R_i=0\) were excluded from the treatment-effect analysis but retained in the full baseline cohort used to estimate the censoring distribution. We estimated \(G(s\mid A_i,X_i,Z_i)\) using five-fold cross-fitting of a Cox proportional hazards model, with \(\widetilde U_i\) as the follow-up time and \(\widetilde{\Delta}_i\) as the censoring-event indicator.

Each participant with \(R_i=1\) was assigned the weight \(\widehat\omega_i=1/\widehat G\{\min(U_i,t_0)\mid A_i,X_i,Z_i\}\). Thus, a participant who developed MCI at month \(t\le t_0\) was weighted by \(1/\widehat G(t\mid A_i,X_i,Z_i)\), whereas a participant known to remain MCI-free through \(t_0\) was weighted by \(1/\widehat G(t_0\mid A_i,X_i,Z_i)\). Under conditional independent censoring and positivity, these weights allow participants with observed restricted outcomes to represent participants with similar baseline characteristics whose follow-up ended before their restricted outcomes could be determined.

Our application was designed to demonstrate Setting~II, in which \(W^\star\) would ideally be estimated from an external study observing both high-risk membership and the predictors available in the analysis cohort. Due to lack of a nationwide external study with sufficiently comprehensive and harmonizable baseline covariates, we created a nonoverlapping internal emulation: the \(83{,}745\) participants with \(R_i=0\) formed the membership-score reference sample and contributed no outcomes to treatment-effect estimation, while the \(37{,}577\) participants with \(R_i=1\) formed the treatment-effect analysis sample.

Within the reference sample, we defined \(H_i=1\) when hypertension, type 2 diabetes, heart failure, or stroke was present at baseline. Let \(V_i\) contain the demographic, nondefining comorbidity, lifestyle, housing, and SDOH predictors, excluding the four defining conditions and all medication indicators. In each completed dataset, we fitted \(W^\star(V_i,Z_i)=\Pr(H_i=1\mid V_i,Z_i)\) by \(L_2\)-penalized logistic regression using only \((V_i,Z_i,H_i)\) from the reference sample, froze the fitted model, and applied it to the analysis sample. The target estimand was
\[
\exp\{\psi(z)\}=\frac{\exp\left[E\{\log(\min[T_i(1),t_0])\mid H_i=1,Z_i=z\}\right]}{\exp\left[E\{\log(\min[T_i(0),t_0])\mid H_i=1,Z_i=z\}\right]},
\]
the ratio of geometric mean 48-month restricted MCI-free time comparing medication use with non-use within high-risk subgroup \(z\); values above one indicate longer restricted MCI-free time under medication use.

Treatment propensities and outcome regressions were estimated in the analysis sample using five-fold cross-fitting with regularized XGBoost classifiers and regressors. The Cox censoring model included \((X_i,Z_i)\) and all three medication indicators.
We used the constructible \(H_i\) only for the FS-HR and HR comparators.

\subsection{Results}

Among \(121{,}322\) baseline participants, \(83{,}745\) formed the reference sample and \(37{,}577\) formed the outcome-analysis sample, which included 500 MCI events within 48 months. High-risk prevalence was similar in the two samples (40.5\% and 43.7\%, respectively). For both medications, RSW confidence intervals were narrower than the oracle hard-targeting intervals in the low-prevalence Clusters~I and~II and similar across methods in Cluster~III.

\begin{figure}[H]
\centering
\includegraphics[width=.98\textwidth,alt={Forest plot of overall and four cluster-specific time ratio estimates for two medication classes. For antihypertensive use, confidence intervals lie above one overall and in Cluster III.}]{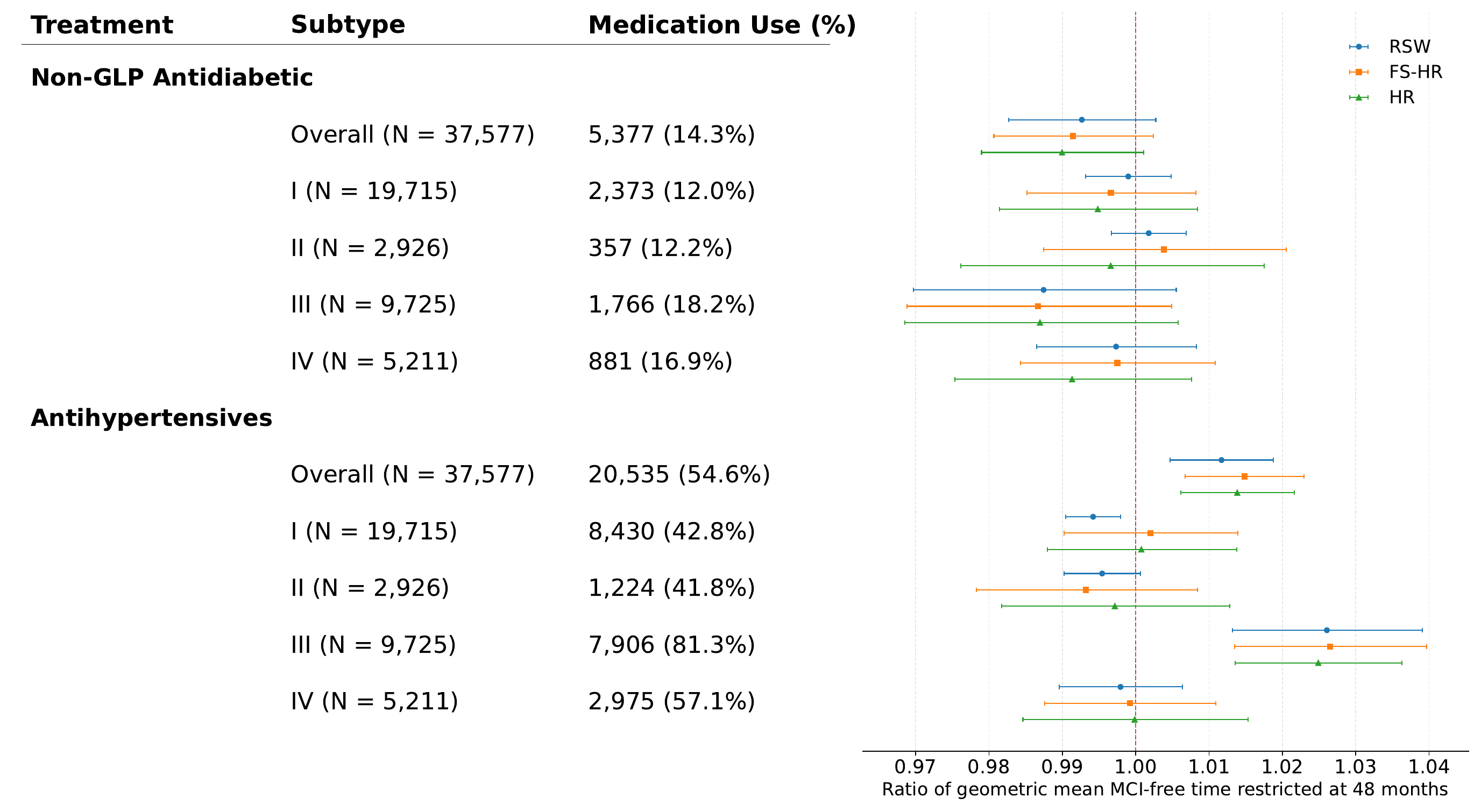}
\caption{Estimated effects of non-GLP antidiabetic and antihypertensive medication use on MCI-free time restricted at 48 months. Points and horizontal bars denote exponentiated treatment-effect estimates and 95\% confidence intervals from the RSW analysis and the oracle FS-HR and HR comparators. Values above one indicate longer restricted MCI-free time among medication users. Medication-use counts and percentages in the full outcome-analysis sample are shown for context.}
\label{fig:journal_forest}
\end{figure}

Non-GLP antidiabetic use showed no clear association with restricted MCI-free time. Antihypertensive use was associated with modestly longer restricted MCI-free time overall (RSW time ratio, 1.012; 95\% CI, 1.005--1.019) and in Cluster~III (1.026; 95\% CI, 1.013--1.039), with similar conclusions from FS-HR and HR (Figure~\ref{fig:journal_forest}).

These effects are geometric-mean time ratios for \(\min(T,48)\), not differences in MCI-free months or conventional arithmetic restricted mean survival times. Their magnitudes were small and should be interpreted cautiously given the limited number of events.

\section{Discussion}

We proposed RSW to estimate subgroup-specific treatment effects in a prespecified high-risk population without requiring known risk status or discarding information from participants not in the high-risk subset. Three findings summarize its empirical behavior. First, RSW preserved small bias while improving precision relative to the two hard-targeting alternatives. Second, this gain was largest when the observed high-risk-by-subgroup cells were small and decreased when almost the entire subgroup was high risk. Third, violation of treatment-effect sufficiency introduced a clear bias--variance trade-off: RSW remained the lowest-MSE estimator in the scenarios examined, but the hard-targeting estimators retained better bias and coverage under strong violations. The effect was modest in the binary-outcome scenarios considered. RSW also relies on accurate estimation and, in Setting~II, transportability of the target-membership probability \(W\); misspecification may bias estimation of the high-risk GATE.

There are several limitations to our application of RSW to the All\textit{of}Us study. We used the non-overlapping internal reference sample to emulate a genuine external sample. Differences in enrollment timing or follow-up could affect risk-score transportability. However, baseline high-risk prevalence was similar in both samples, which partially mitigates the risk of non-transportability. Flexible nuisance modeling and cross-fitting improve robustness to model specification but cannot address violations caused by unmeasured determinants of follow-up. Additional limitations arise from the observational exposure and outcome definitions: medication use was measured only at baseline without information on initiation or adherence, and the endpoint was the first recorded MCI diagnosis in the EHR rather than biological onset. A well-designed prospective study may provide data less subject to these limitations. Although here we focus on GATEs with a low-dimensional subgroup variable $Z$, the framework extends naturally to conditional average treatment effects by replacing $Z$ with a vector of treatment-effect modifiers and estimating the corresponding treatment-effect surface jointly with $W$.

\section*{Acknowledgments}
This research is supported by U.S. NIH grants NS073671, GM124104, and AG097722.

\bibliographystyle{plainnat}
\bibliography{latex/refs}

@article{Zhao2025.05.27.25328416,
  author  = {Zhao, Yinjun and Wang, Yuanjia and Liu, Ying},
  title   = {An Integrative Approach for Subtyping Mental Disorders Using Multimodal Data},
  journal = {medRxiv.},
  year    = {2025},
  note    = { doi:10.1101/2025.05.27.25328416}
}

@article{FrangakisRubin2002,
  author  = {Frangakis, Constantine E. and Rubin, Donald B.},
  title   = {Principal Stratification in Causal Inference},
  journal = {Biometrics},
  year    = {2002},
  volume  = {58},
  number  = {1},
  pages   = {21--29}
}

@article{DingLu2017,
  author  = {Ding, Peng and Lu, Jiannan},
  title   = {Principal Stratification Analysis Using Principal Scores},
  journal = {Journal of the Royal Statistical Society: Series B},
  year    = {2017},
  volume  = {79},
  number  = {3},
  pages   = {757--777}
}

@article{AllofUs2019,
  author = {{All of Us Research Program Investigators}},
  title = {The ``{All of Us}'' Research Program},
  journal = {New England Journal of Medicine},
  year = {2019},
  volume = {381},
  number = {7},
  pages = {668--676}
}

@article{Livingston2024,
  author = {Livingston, Gill and Huntley, Jonathan and Liu, Kathy Y. and Costafreda, Sergi G. and Selb{\ae}k, Geir and Alladi, Suvarna and others},
  title = {Dementia Prevention, Intervention, and Care: 2024 Report of the {Lancet} Standing Commission},
  journal = {The Lancet},
  year = {2024},
  volume = {404},
  number = {10452},
  pages = {572--628}
}

@article{Williamson2019,
  author = {Williamson, Jeff D. and Pajewski, Nicholas M. and Auchus, Alexander P. and Bryan, R. Nick and Chelune, Gordon and Cheung, Alfred K. and others},
  title = {Effect of Intensive vs Standard Blood Pressure Control on Probable Dementia: A Randomized Clinical Trial},
  journal = {JAMA},
  year = {2019},
  volume = {321},
  number = {6},
  pages = {553--561}
}

@article{Ngandu2015FINGER,
  author = {Ngandu, Tiia and Lehtisalo, Jenni and Solomon, Alina and Lev{\"a}lahti, Esko and Ahtiluoto, Satu and Antikainen, Riitta and others},
  title = {A 2 year multidomain intervention of diet, exercise, cognitive training, and vascular risk monitoring versus control to prevent cognitive decline in at-risk elderly people ({FINGER}): a randomised controlled trial},
  journal = {The Lancet},
  year = {2015},
  volume = {385},
  number = {9984},
  pages = {2255--2263},
  doi = {10.1016/S0140-6736(15)60461-5}
}

@article{Solomon2018FINGERAPOE,
  author = {Solomon, Alina and Turunen, Heidi and Ngandu, Tiia and Peltonen, Markku and Lev{\"a}lahti, Esko and Helisalmi, Seppo and others},
  title = {Effect of the Apolipoprotein {E} Genotype on Cognitive Change During a Multidomain Lifestyle Intervention: A Subgroup Analysis of a Randomized Clinical Trial},
  journal = {JAMA Neurology},
  year = {2018},
  volume = {75},
  number = {4},
  pages = {462--470},
  doi = {10.1001/jamaneurol.2017.4365}
}

@article{Delrieu2019MAPTAmyloid,
  author = {Delrieu, Julien and Payoux, Pierre and Carri{\'e}, Isabelle and Cantet, Christelle and Weiner, Michael and Vellas, Bruno and others},
  title = {Multidomain Intervention and/or Omega-3 in Nondemented Elderly Subjects According to Amyloid Status},
  journal = {Alzheimer's \& Dementia},
  year = {2019},
  volume = {15},
  number = {11},
  pages = {1392--1401},
  doi = {10.1016/j.jalz.2019.07.008}
}

@article{Ghazi2023SPRINTRisk,
  author = {Ghazi, Lama and Shen, Jincheng and Ying, Jian and Derington, Catherine G. and Cohen, Jordana B. and Marcum, Zachary A. and others},
  title = {Identifying Patients for Intensive Blood Pressure Treatment Based on Cognitive Benefit: A Secondary Analysis of the {SPRINT} Randomized Clinical Trial},
  journal = {JAMA Network Open},
  year = {2023},
  volume = {6},
  number = {5},
  pages = {e2314443},
  doi = {10.1001/jamanetworkopen.2023.14443}
}

@article{Brookmeyer2019PreventionTrialDesign,
  author = {Brookmeyer, Ron and Abdalla, Nada and Kawas, Claudia H. and Corrada, Mar{\'i}a M.},
  title = {Design and Sample Size Considerations for {Alzheimer's} Disease Prevention Trials Using Clinical Outcomes},
  journal = {Alzheimer's \& Dementia},
  year = {2019},
  volume = {15},
  number = {5},
  pages = {593--600},
  doi = {10.1016/j.jalz.2018.12.014}
}

@article{Duara2022ADHeterogeneity,
  author  = {Duara, Ranjan and Barker, Warren W.},
  title   = {Heterogeneity in {Alzheimer's Disease} Diagnosis and Progression Rates: Implications for Therapeutic Trials},
  journal = {Neurotherapeutics},
  year    = {2022},
  volume  = {19},
  number  = {1},
  pages   = {8--25},
  doi     = {10.1007/s13311-022-01185-z},
  pmid    = {35084721},
  pmcid   = {PMC9130395}
}

@article{Jack2018NIAAAFramework,
  title   = {{NIA-AA} Research Framework: Toward a biological definition of {Alzheimer's} disease},
  author  = {Jack, Clifford R. and Bennett, David A. and Blennow, Kaj and Carrillo, Maria C. and Dunn, Billy and Haeberlein, Samantha Budd and others},
  journal = {Alzheimer's \& Dementia},
  year    = {2018},
  volume  = {14},
  number  = {4},
  pages   = {535--562},
  doi     = {10.1016/j.jalz.2018.02.018}
}

@article{Soldan2016PreclinicalGroups,
  title   = {Hypothetical preclinical {Alzheimer} disease groups and longitudinal cognitive change},
  author  = {Soldan, Anja and Pettigrew, Corinne and Cai, Qing and Wang, Mei-Cheng and Moghekar, Abhay R. and O'Brien, Richard J. and others},
  journal = {JAMA Neurology},
  year    = {2016},
  volume  = {73},
  number  = {6},
  pages   = {698--705},
  doi     = {10.1001/jamaneurol.2016.0194}
}

@article{chernozhukov2018double,
  author  = {Chernozhukov, Victor and Chetverikov, Denis and Demirer, Mert and Duflo, Esther and Hansen, Christian and Newey, Whitney and others},
  title   = {Double/Debiased Machine Learning for Treatment and Structural Parameters},
  journal = {The Econometrics Journal},
  year    = {2018},
  volume  = {21},
  number  = {1},
  pages   = {C1--C68},
  doi     = {10.1111/ectj.12097}
}

@article{wager2018causal,
  author  = {Wager, Stefan and Athey, Susan},
  title   = {Estimation and Inference of Heterogeneous Treatment Effects Using Random Forests},
  journal = {Journal of the American Statistical Association},
  year    = {2018},
  volume  = {113},
  number  = {523},
  pages   = {1228--1242},
  doi     = {10.1080/01621459.2017.1319839}
}

@article{brookmeyer2019design,
  author  = {Brookmeyer, Ron and Abdalla, Nada},
  title   = {Design and Sample Size Considerations for Alzheimer's Disease Prevention Trials Using Multistate Models},
  journal = {Clinical Trials},
  year    = {2019},
  volume  = {16},
  number  = {2},
  pages   = {111--119},
  doi     = {10.1177/1740774518816323}
}

@book{ImbensRubin2015,
  author    = {Imbens, Guido W. and Rubin, Donald B.},
  title     = {Causal Inference for Statistics, Social, and Biomedical Sciences: An Introduction},
  publisher = {Cambridge University Press},
  year      = {2015},
  address   = {Cambridge}
}

@book{HernanRobins2020,
  author    = {Hern{\'a}n, Miguel A. and Robins, James M.},
  title     = {Causal Inference: What If},
  publisher = {Chapman \& Hall/CRC},
  year      = {2020},
  address   = {Boca Raton}
}

@article{RobinsRotnitzky1995,
  author  = {Robins, James M. and Rotnitzky, Andrea},
  title   = {Semiparametric Efficiency in Multivariate Regression Models with Missing Data},
  journal = {Journal of the American Statistical Association},
  year    = {1995},
  volume  = {90},
  number  = {429},
  pages   = {122--129},
  doi     = {10.1080/01621459.1995.10476494}
}

@article{SemenovaChernozhukov2021,
  author  = {Semenova, Vira and Chernozhukov, Victor},
  title   = {Debiased Machine Learning of Conditional Average Treatment Effects and Other Causal Functions},
  journal = {The Econometrics Journal},
  year    = {2021},
  volume  = {24},
  number  = {2},
  pages   = {264--289},
  doi     = {10.1093/ectj/utaa027}
}

@article{Wang2021SubgroupStatistical,
  author = {Wang, Xi and Piantadosi, Steven and Le-Rademacher, Jennifer and Mandrekar, Sumithra J.},
  title = {Statistical Considerations for Subgroup Analyses},
  journal = {Journal of Thoracic Oncology},
  year = {2021},
  volume = {16},
  number = {3},
  pages = {375--380},
  doi = {10.1016/j.jtho.2020.12.008}
}

@article{Coley2022RCTReview,
  author = {Coley, Nicola and Giulioli, Caroline and Aisen, Paul S. and Vellas, Bruno and Andrieu, Sandrine},
  title = {Randomised controlled trials for the prevention of cognitive decline or dementia: A systematic review},
  journal = {Ageing Research Reviews},
  year = {2022},
  volume = {82},
  pages = {101777},
  doi = {10.1016/j.arr.2022.101777}
}

@article{Andrieu2017MAPT,
  author = {Andrieu, Sandrine and Guyonnet, Sophie and Coley, Nicola and Cantet, Christelle and Bonnefoy, Marc and Bordes, Serge and others},
  title = {Effect of long-term omega 3 polyunsaturated fatty acid supplementation with or without multidomain intervention on cognitive function in elderly adults with memory complaints ({MAPT}): A randomised, placebo-controlled trial},
  journal = {The Lancet Neurology},
  year = {2017},
  volume = {16},
  number = {5},
  pages = {377--389},
  doi = {10.1016/S1474-4422(17)30040-6}
}

@article{chernozhukov2025generic,
  title={{Fisher--Schultz} Lecture: Generic Machine Learning Inference on Heterogeneous Treatment Effects in Randomized Experiments, With an Application to Immunization in {India}},
  author={Chernozhukov, Victor and Demirer, Mert and Duflo, Esther and Fern{\'a}ndez-Val, Iv{\'a}n},
  journal={Econometrica},
  volume={93},
  number={4},
  pages={1121--1164},
  year={2025},
  doi={10.3982/ECTA19303}
}

@article{imai2025statistical,
  title={Statistical Inference for Heterogeneous Treatment Effects Discovered by Generic Machine Learning in Randomized Experiments},
  author={Imai, Kosuke and Li, Michael Lingzhi},
  journal={Journal of Business \& Economic Statistics},
  volume={43},
  number={1},
  pages={256--268},
  year={2025},
  doi={10.1080/07350015.2024.2358909}
}

@article{park2024groupwise,
  title={A Groupwise Approach for Inferring Heterogeneous Treatment Effects in Causal Inference},
  author={Park, Chan and Kang, Hyunseung},
  journal={Journal of the Royal Statistical Society Series A: Statistics in Society},
  volume={187},
  number={2},
  pages={374--392},
  year={2024},
  doi={10.1093/jrsssa/qnad125}
}

@article{kennedy2023towards,
  title={Towards Optimal Doubly Robust Estimation of Heterogeneous Causal Effects},
  author={Kennedy, Edward H.},
  journal={Electronic Journal of Statistics},
  volume={17},
  number={2},
  pages={3008--3049},
  year={2023},
  doi={10.1214/23-EJS2157}
}

@article{robinson1988root,
  title={Root-N-Consistent Semiparametric Regression},
  author={Robinson, Peter M.},
  journal={Econometrica},
  volume={56},
  number={4},
  pages={931--954},
  year={1988},
  doi={10.2307/1912705}
}

@article{RobinsFinkelstein2000,
  author  = {Robins, James M. and Finkelstein, Dianne M.},
  title   = {Correcting for Noncompliance and Dependent Censoring in an {AIDS}
             Clinical Trial with Inverse Probability of Censoring Weighted
             ({IPCW}) Log-Rank Tests},
  journal = {Biometrics},
  year    = {2000},
  volume  = {56},
  number  = {3},
  pages   = {779--788},
  doi     = {10.1111/j.0006-341X.2000.00779.x}
}

@article{Albert2011MCI,
  author  = {Albert, Marilyn S. and DeKosky, Steven T. and Dickson, Dennis
             and Dubois, Bruno and Feldman, Howard H. and Fox, Nick C. and others},
  title   = {The Diagnosis of Mild Cognitive Impairment Due to {Alzheimer's}
             Disease: Recommendations from the {National Institute on
             Aging--Alzheimer's Association} Workgroups on Diagnostic
             Guidelines for {Alzheimer's} Disease},
  journal = {Alzheimer's \& Dementia},
  year    = {2011},
  volume  = {7},
  number  = {3},
  pages   = {270--279},
  doi     = {10.1016/j.jalz.2011.03.008}
}

@article{CukiermanYaffe2020REWIND,
  author  = {Cukierman-Yaffe, Tali and Gerstein, Hertzel C.
             and Colhoun, Helen M. and Diaz, Rafael and Garc{\'i}a-P{\'e}rez, Luis-Emilio
             and Lakshmanan, Mark and others},
  title   = {Effect of Dulaglutide on Cognitive Impairment in Type 2
             Diabetes: An Exploratory Analysis of the {REWIND} Trial},
  journal = {The Lancet Neurology},
  year    = {2020},
  volume  = {19},
  number  = {7},
  pages   = {582--590},
  doi     = {10.1016/S1474-4422(20)30173-3}
}

@book{Rubin1987,
  author    = {Rubin, Donald B.},
  title     = {Multiple Imputation for Nonresponse in Surveys},
  publisher = {John Wiley \& Sons},
  address   = {New York},
  year      = {1987},
  doi       = {10.1002/9780470316696},
  isbn      = {9780471087052}
}

@article{Newey1994,
  author  = {Newey, Whitney K.},
  title   = {The Asymptotic Variance of Semiparametric Estimators},
  journal = {Econometrica},
  year    = {1994},
  volume  = {62},
  number  = {6},
  pages   = {1349--1382},
  doi     = {10.2307/2951752}
}

\section*{Supplementary Material}
Additional proofs and simulation tables are provided in the Supplementary Material accompanying this preprint.

\end{document}

\begin{center}
\vspace*{1em}
{\LARGE\bfseries
Supplementary Material for\\[0.35em]
``Estimating Group Average Treatment Effects in High-Risk Populations with Unobserved Target Membership''}

\vspace{1.5em}

{\large\bfseries Shubo Zhang\(^{1}\), Yinjun Zhao\(^{1}\), Donglin Zeng\(^{2}\), and Yuanjia Wang\(^{1,*}\)\par}

\vspace{0.9em}

\(^{1}\)Department of Biostatistics, Mailman School of Public Health, Columbia University\\
\(^{2}\)Department of Biostatistics, School of Public Health, University of Michigan\\
\(^{*}\)\textit{email:} yw2016@cumc.columbia.edu
\end{center}

\vspace{2em}
\section{Theoretical results}
\label{sec:supp-theory}
\subsection{Risk-weighted representation}
\label{app:weighted}
For any \(z\) with \(P(H_i=1,Z_i=z)>0\),
\begin{align}
\psi(z)&=\frac{E[\{Y_i(1)-Y_i(0)\}H_i\mid Z_i=z]}{P(H_i=1\mid Z_i=z)}. \label{eq:supp_target_ratio}
\end{align}
Assumption~(A4) gives \(E\{Y_i(1)-Y_i(0)\mid H_i=1,W_i,Z_i\}=\tau(Z_i,W_i)\). In addition, since \(W_i=E(H_i\mid X_i,Z_i)\), iterated expectations imply
\begin{equation}
E(H_i\mid W_i,Z_i)=E\{E(H_i\mid X_i,Z_i)\mid W_i,Z_i\}=W_i. \label{eq:supp_balancing}
\end{equation}
Substituting \eqref{eq:supp_balancing} into \eqref{eq:supp_target_ratio} yields
\begin{equation}
\psi(z)=\frac{E\{\tau(z,W_i)W_i\mid Z_i=z\}}{E(W_i\mid Z_i=z)}, \label{eq:supp_weighted_representation}
\end{equation}
which proves the risk-weighted representation stated in the main text.
\subsection{Population estimating equations}
\label{app:population_moments}
Let \(\pi_0\) and \(\mu_{a,0}\), \(a\in\{0,1\}\), denote the true treatment propensity and outcome regressions defined in the main text:
\begin{equation*}
\begin{aligned}
\pi_0(x,z,w)&=P(A_i=1\mid X_i=x,Z_i=z,W_i=w),\\
\mu_{a,0}(x,z,w)&=E(Y_i\mid A_i=a,X_i=x,Z_i=z,W_i=w).
\end{aligned}
\end{equation*}
Let \(\Gamma_i=\Gamma_i(\eta_0)\) denote the oracle doubly robust signal defined in the main text. Under (A1)--(A3),
\begin{equation*}
E(\Gamma_i\mid X_i,Z_i)=E\{Y_i(1)-Y_i(0)\mid X_i,Z_i\},\qquad E(\Gamma_i\mid Z_i,W_i)=\tau(Z_i,W_i).
\end{equation*}
Let
\begin{equation*}
Q=E\{b(Z_i,W_i)b(Z_i,W_i)^\top\},\qquad \beta_0=Q^{-1}E\{b(Z_i,W_i)\Gamma_i\}.
\end{equation*}
Then
\begin{equation*}
E[b(Z_i,W_i)\{\Gamma_i-b(Z_i,W_i)^\top\beta_0\}]=0,
\end{equation*}
Under the target-compatibility condition B5 below, the population equation for the original high-risk subgroup effect is
\begin{equation}
E[\mathbb I(Z_i=z)W_i\{b(z,W_i)^\top\beta_0-\psi(z)\}]=0. \label{eq:supp_psi_moment}
\end{equation}
\subsection{Inference for the hard-targeting benchmarks}
\label{app:hard_inference}
FS-HR and HR are standard groupwise DML estimators because their final targets are defined by the prespecified cells \(\{H_i=1,Z_i=z\}\), rather than by estimated risk-score weights. Under positive cell probability, treatment overlap, conditional exchangeability, and the usual cross-fitted nuisance-rate conditions, their asymptotic normality and orthogonal-score variance estimators follow directly from existing groupwise DML results \citep{park2024groupwise}.
\subsection{Regularity conditions}
\label{app:rsw_regularity}
\paragraph{B1 (sampling, moments, and overlap)} The observations are i.i.d., the target subgroup has positive probability, and \(E|Y_i|^q<\infty\) and \(E\|r(X_i,Z_i)\|^q<\infty\) for some \(q>4\). In addition,
\begin{equation*}
P(H_i=1,Z_i=z)>0,\qquad \epsilon_A\le\pi_0(X_i,Z_i,W_i)\le1-\epsilon_A\ \text{a.s.},\qquad 0<\sigma^2(z)<\infty.
\end{equation*}
\paragraph{B2 (cross-fitted nuisance rates)} For each fold, the nuisance learners and every generated feature used to train them are constructed only from that fold's training observations. The estimated propensity scores remain uniformly bounded away from zero and one, and, uniformly over the fixed number of folds,
\begin{align}
\max_{a\in\{0,1\}}\|\widehat\mu_a-\mu_{a,0}\|_{P,2}&=o_p(1),\notag\\
\|\widehat\pi-\pi_0\|_{P,2}&=o_p(1),\notag\\
\max_{a\in\{0,1\}}\|\widehat\mu_a-\mu_{a,0}\|_{P,2}\|\widehat\pi-\pi_0\|_{P,2}&=o_p(n^{-1/2}). \label{eq:supp_nuisance_rates}
\end{align}
\paragraph{B3 (membership-score model)} The correctly specified model
\begin{equation*}
W(X_i,Z_i;\alpha)=\operatorname{expit}\{r(X_i,Z_i)^\top\alpha\}
\end{equation*}
has a fixed-dimensional, full-rank feature vector and a unique true coefficient \(\alpha_0\). The matrix \(E\{W_i(1-W_i)r_i r_i^\top\}\) is positive definite, and each training-fold estimator satisfies
\begin{align}
\widehat\alpha^{(-k)}-\alpha_0&=\frac{1}{|\mathcal I_k^c|}\sum_{j\in\mathcal I_k^c}\mathrm{IF}_{\alpha}(O_j)+o_p(n^{-1/2}),\notag\\
\mathrm{IF}_{\alpha}(O_i)&=\left[E\{W_i(1-W_i)r_i r_i^\top\}\right]^{-1}r_i(H_i-W_i), \label{eq:supp_alpha_expansion}
\end{align}
where \(r_i=r(X_i,Z_i)\). If an \(L_2\) penalty is used in an average-loss objective, its effective coefficient is required to be \(o(n^{-1/2})\).
\paragraph{B4 (basis regularity)} The dimension of \(b(z,w)\) is fixed; \(b(z,w)\), \(\partial_w b(z,w)\), and \(\partial_w^2b(z,w)\) are uniformly bounded; and the eigenvalues of \(Q\) are bounded away from zero.
\paragraph{B5 (target-compatible basis)} For each target subgroup \(z\), the population basis projection preserves the risk-weighted treatment-effect target:
\begin{equation*}
E\left[\mathbb I(Z_i=z)W_i\left\{b(z,W_i)^\top\beta_0-\tau(z,W_i)\right\}\right]=0.
\end{equation*}
Together with the risk-weighted representation in \eqref{eq:supp_weighted_representation}, this condition implies the population moment in \eqref{eq:supp_psi_moment} for \(\psi(z)\).
\subsection{Influence function under Setting I}
\label{app:rsw_if}
The following stacked-moment correction accounts for the generated membership score and follows standard two-step semiparametric influence-function calculations \citep{Newey1994}.
Write \(b_i=b(Z_i,W_i)\), \(\dot b_i=\partial_wb(Z_i,W_i)\), and \(\dot W_i=W_i(1-W_i)r_i\). Define
\begin{align*}
\phi_{\alpha,i}&=r_i(H_i-W_i),\\
\phi_{\beta,i}&=b_i(\Gamma_i-b_i^\top\beta_0),\\
\phi_{\psi,i}(z)&=\mathbb I(Z_i=z)W_i\{b(z,W_i)^\top\beta_0-\psi(z)\}.
\end{align*}
The nonzero blocks of the lower-triangular Jacobian are
\begin{align}
A_{11}&=-E\{W_i(1-W_i)r_i r_i^\top\},\notag\\
A_{22}&=-E(b_i b_i^\top),\notag\\
A_{21}&=E[\{\dot b_i(\Gamma_i-b_i^\top\beta_0)-b_i\dot b_i^\top\beta_0\}\dot W_i^\top], \label{eq:A21}\\
A_{33}(z)&=-E\{\mathbb I(Z_i=z)W_i\},\\
A_{32}(z)&=E\{\mathbb I(Z_i=z)W_i b(z,W_i)^\top\},\\
A_{31}(z)&=E[\mathbb I(Z_i=z)\{b(z,W_i)^\top\beta_0-\psi(z)+W_i\partial_wb(z,W_i)^\top\beta_0\}\dot W_i^\top]. \label{eq:A31}
\end{align}
Consequently,
\begin{align*}
\mathrm{IF}_{\alpha}(O_i)&=-A_{11}^{-1}\phi_{\alpha,i},\\
\mathrm{IF}_{\beta}(O_i)&=-A_{22}^{-1}\{\phi_{\beta,i}+A_{21}\mathrm{IF}_{\alpha}(O_i)\},\\
\mathrm{IF}_{\psi,\textnormal{I}}(O_i;z)&=-A_{33}(z)^{-1}\{\phi_{\psi,i}(z)+A_{31}(z)\mathrm{IF}_{\alpha}(O_i)+A_{32}(z)\mathrm{IF}_{\beta}(O_i)\}.
\end{align*}
\subsection{Proof of the Setting I inference theorem}
\label{app:rsw_proof}
\paragraph{Step 1: Consistency} Conditions B2--B4 imply \(\mathbb P_n\|b(Z_i,\widehat W_i)-b_i\|^2=o_p(1)\) and \(\|\widehat\Gamma_i-\Gamma_i\|_{P,2}=o_p(1)\). Conditional laws of large numbers for the fixed number of folds therefore yield
\begin{equation*}
\mathbb P_n\{b(Z_i,\widehat W_i)b(Z_i,\widehat W_i)^\top\}\to_p Q,\qquad \mathbb P_n\{b(Z_i,\widehat W_i)\widehat\Gamma_i\}\to_p E(b_i\Gamma_i).
\end{equation*}
It follows that \(\widehat\beta\to_p\beta_0\). Applying the same argument to the numerator and denominator of \(\widehat\psi_{\mathrm{RSW}}(z)\), together with B5 and \eqref{eq:supp_psi_moment}, gives \(\widehat\psi_{\mathrm{RSW}}(z)\to_p\psi(z)\).
\paragraph{Step 2: Orthogonal-signal remainder} For validation observations in fold \(k\), direct calculation conditional on the training sample gives
\begin{equation}
E(\widehat\Gamma_i-\Gamma_i\mid X_i,Z_i,\mathcal I_k^c)=(\widehat\pi_i-\pi_{0,i})\left\{\frac{\widehat\mu_{1,i}-\mu_{1,0,i}}{\widehat\pi_i}+\frac{\widehat\mu_{0,i}-\mu_{0,0,i}}{1-\widehat\pi_i}\right\}. \label{eq:supp_dr_bias}
\end{equation}
By overlap, Cauchy--Schwarz, and \eqref{eq:supp_nuisance_rates}, the expectation of \eqref{eq:supp_dr_bias} multiplied by any bounded basis component is \(o_p(n^{-1/2})\). Cross-fitting and conditional Chebyshev's inequality control the centered empirical term, so
\begin{equation}
\sqrt n\,\mathbb P_n[b(Z_i,\widehat W_i)(\widehat\Gamma_i-\Gamma_i)]=o_p(1). \label{eq:supp_dml_remainder}
\end{equation}
This is the usual cross-fitted DML product-rate remainder \citep{chernozhukov2018double,kennedy2023towards}.
\paragraph{Step 3: Cross-fitted membership-score expansion} For any twice continuously differentiable vector function \(f(O_i;\alpha)\), a foldwise Taylor expansion gives
\begin{align}
\sum_{k=1}^K\frac{|\mathcal I_k|}{n}\mathbb P_{n,k}f(O_i;\widehat\alpha^{(-k)})&=\mathbb P_nf(O_i;\alpha_0)\nonumber\\
&\quad+E\{\partial_\alpha f(O_i;\alpha_0)\}\sum_{k=1}^K\frac{|\mathcal I_k|}{n}(\widehat\alpha^{(-k)}-\alpha_0)\nonumber\\
&\quad+o_p(n^{-1/2}). \label{eq:supp_fold_taylor}
\end{align}
Because every observation belongs to \(K-1\) training samples and the folds are balanced, \eqref{eq:supp_alpha_expansion} implies
\begin{equation*}
\sum_{k=1}^K\frac{|\mathcal I_k|}{n}(\widehat\alpha^{(-k)}-\alpha_0)=\mathbb P_n\mathrm{IF}_{\alpha}(O_i)+o_p(n^{-1/2}).
\end{equation*}
Thus cross-fitting removes own-observation overfitting but does not remove the first-order generated-regressor contribution.
\paragraph{Step 4: Linearization} The OLS equation, \eqref{eq:supp_dml_remainder}, and the expansion in \eqref{eq:supp_fold_taylor} give
\begin{equation*}
\sqrt n(\widehat\beta-\beta_0)=\frac{1}{\sqrt n}\sum_{i=1}^n\mathrm{IF}_{\beta}(O_i)+o_p(1).
\end{equation*}
Applying the same expansion to the weighted equation
\begin{equation*}
\mathbb P_n[\mathbb I(Z_i=z)\widehat W_i\{b(z,\widehat W_i)^\top\widehat\beta-\widehat\psi_{\mathrm{RSW}}(z)\}]=0
\end{equation*}
yields
\begin{equation*}
\sqrt n\{\widehat\psi_{\mathrm{RSW}}(z)-\psi(z)\}=\frac{1}{\sqrt n}\sum_{i=1}^n\mathrm{IF}_{\psi,\textnormal{I}}(O_i;z)+o_p(1).
\end{equation*}
The multivariate central limit theorem establishes asymptotic normality.
\subsection{Sample analogues and variance consistency}
\label{app:rsw_variance}
Let \(\widehat b_i=b(Z_i,\widehat W_i)\), \(\widehat{\dot b}_i=\partial_wb(Z_i,\widehat W_i)\), and \(\widehat{\dot W}_i=\widehat W_i(1-\widehat W_i)r_i\). The sample Jacobian blocks are
\begin{align*}
\widehat A_{11}&=-\mathbb P_n\{\widehat W_i(1-\widehat W_i)r_i r_i^\top\},\\
\widehat A_{22}&=-\mathbb P_n(\widehat b_i\widehat b_i^\top),\\
\widehat A_{21}&=\mathbb P_n[\{\widehat{\dot b}_i(\widehat\Gamma_i-\widehat b_i^\top\widehat\beta)-\widehat b_i\widehat{\dot b}_i^\top\widehat\beta\}\widehat{\dot W}_i^\top],\\
\widehat A_{33}(z)&=-\mathbb P_n\{\mathbb I(Z_i=z)\widehat W_i\},\\
\widehat A_{32}(z)&=\mathbb P_n\{\mathbb I(Z_i=z)\widehat W_i b(z,\widehat W_i)^\top\},\\
\widehat A_{31}(z)&=\mathbb P_n\left[\mathbb I(Z_i=z)\left\{b(z,\widehat W_i)^\top\widehat\beta-\widehat\psi_{\mathrm{RSW}}(z)\right.\right.\\
&\hspace{42mm}\left.\left.+\widehat W_i\partial_wb(z,\widehat W_i)^\top\widehat\beta\right\}\widehat{\dot W}_i^\top\right].
\end{align*}
Define
\begin{align*}
\widehat\phi_{\alpha,i}&=r_i(H_i-\widehat W_i),\\
\widehat\phi_{\beta,i}&=\widehat b_i(\widehat\Gamma_i-\widehat b_i^\top\widehat\beta),\\
\widehat\phi_{\psi,i}(z)&=\mathbb I(Z_i=z)\widehat W_i\{b(z,\widehat W_i)^\top\widehat\beta-\widehat\psi_{\mathrm{RSW}}(z)\}.
\end{align*}
The plug-in influence functions are
\begin{align*}
\widehat{\mathrm{IF}}_{\alpha,i}&=-\widehat A_{11}^{-1}\widehat\phi_{\alpha,i},\\
\widehat{\mathrm{IF}}_{\beta,i}&=-\widehat A_{22}^{-1}\{\widehat\phi_{\beta,i}+\widehat A_{21}\widehat{\mathrm{IF}}_{\alpha,i}\},\\
\widehat{\mathrm{IF}}_{\psi,\textnormal{I},i}(z)&=-\widehat A_{33}(z)^{-1}\{\widehat\phi_{\psi,i}(z)+\widehat A_{31}(z)\widehat{\mathrm{IF}}_{\alpha,i}+\widehat A_{32}(z)\widehat{\mathrm{IF}}_{\beta,i}\}.
\end{align*}
Conditions B1--B5 and the preceding expansions imply
\begin{equation*}
\mathbb P_n[\{\widehat{\mathrm{IF}}_{\psi,\textnormal{I},i}(z)-\mathrm{IF}_{\psi,\textnormal{I}}(O_i;z)\}^2]=o_p(1).
\end{equation*}
Therefore
\begin{equation*}
\widehat V_{\mathrm{RSW},\textnormal{I}}(z)
=\mathbb P_n\{\widehat{\mathrm{IF}}_{\psi,\textnormal{I},i}(z)^2\}
\to_p E\{\mathrm{IF}_{\psi,\textnormal{I}}(O_i;z)^2\}
=V_{\mathrm{RSW},\textnormal{I}}(z),
\end{equation*}
which proves variance consistency and completes the proof. \hfill\(\square\)

\subsection{Proof of the Setting II inference theorem}
\label{app:two_sample_if}
Suppose the membership-score coefficient \(\alpha_0\) is estimated in an independent reference sample \(\mathcal R_m=\{O_j=(H_j,X_j,Z_j):j=1,\ldots,m\}\), whereas the treatment-effect projection and RSW aggregation are estimated in an analysis sample \(\mathcal A_n=\{O_i=(Y_i,A_i,X_i,Z_i):i=1,\ldots,n\}\). The two samples are nonoverlapping, and the analysis-sample hard membership labels are not used by RSW. For \(s\in\{A,R\}\), define \(W_s(x,z)=P_s(H=1\mid X=x,Z=z)\), and assume
\begin{equation*}
W_R(x,z)=W_A(x,z)\quad\text{for all }(x,z)\in\operatorname{supp}_A(X,Z).
\end{equation*}
For \(j\in\mathcal R_m\), write \(W_j^R=W_R(X_j,Z_j)\) and \(r_j=r(X_j,Z_j)\), and define
\begin{equation*}
\phi_{\alpha,j}^R=r_j(H_j-W_j^R),\qquad
A_{11}^R=-E_R\{W_j^R(1-W_j^R)r_jr_j^\top\},\qquad
\mathrm{IF}_{\alpha}^R(O_j)=-(A_{11}^R)^{-1}\phi_{\alpha,j}^R.
\end{equation*}
The blocks \(A_{21}\), \(A_{22}\), \(A_{31}(z)\), \(A_{32}(z)\), and \(A_{33}(z)\) retain the definitions in Equations~\eqref{eq:A21}--\eqref{eq:A31}, but their expectations are now taken over the analysis population with \(W_i=W_A(X_i,Z_i)\).
For this two-sample result, conditions B1, B2, B4, and B5 apply to the analysis population. The reference-sample analogue of B3 requires the reference-sample score model to be correctly specified, its information matrix to be nonsingular, and its estimator to admit the displayed asymptotically linear expansion at rate \(m^{-1/2}\).

For \(i\in\mathcal A_n\), write \(W_i=W_A(X_i,Z_i)\) and \(b_i=b(Z_i,W_i)\), and define
\begin{equation*}
\phi_{\beta,i}^A=b_i(\Gamma_i-b_i^\top\beta_0),\qquad
\phi_{\psi,i}^A(z)=\mathbb I(Z_i=z)W_i\{b(z,W_i)^\top\beta_0-\psi(z)\}.
\end{equation*}
Conditional on \(\alpha_0\), the analysis-sample influence functions for the projection and final aggregation are
\begin{align}
\mathrm{IF}_{\beta}^A(O_i)
&=-A_{22}^{-1}\phi_{\beta,i}^A, \label{eq:if_beta_analysis}\\
\mathrm{IF}_{\psi,\textnormal{II}}^A(O_i;z)
&=-A_{33}(z)^{-1}\{\phi_{\psi,i}^A(z)+A_{32}(z)\mathrm{IF}_{\beta}^A(O_i)\}. \label{eq:if_psi_analysis}
\end{align}
The reference-sample perturbation of \(\alpha\) propagates through both downstream stages:
\begin{align*}
\mathrm{IF}_{\beta}^R(O_j)
&=-A_{22}^{-1}A_{21}\mathrm{IF}_{\alpha}^R(O_j),\\
\mathrm{IF}_{\psi,\textnormal{II}}^R(O_j;z)
&=-A_{33}(z)^{-1}\{A_{31}(z)\mathrm{IF}_{\alpha}^R(O_j)
+A_{32}(z)\mathrm{IF}_{\beta}^R(O_j)\} \\
&=-A_{33}(z)^{-1}\{A_{31}(z)-A_{32}(z)A_{22}^{-1}A_{21}\}
\mathrm{IF}_{\alpha}^R(O_j).
\end{align*}
Consequently,
\begin{equation*}
\widehat\psi_{\mathrm{RSW}}(z)-\psi(z)
=\frac{1}{n}\sum_{i=1}^n\mathrm{IF}_{\psi,\textnormal{II}}^A(O_i;z)
+\frac{1}{m}\sum_{j=1}^m\mathrm{IF}_{\psi,\textnormal{II}}^R(O_j;z)
+o_p(n^{-1/2}+m^{-1/2}).
\end{equation*}
Define
\begin{equation*}
V_{\mathrm{RSW},\textnormal{II}}^A(z)=E_A\{\mathrm{IF}_{\psi,\textnormal{II}}^A(O_i;z)^2\},\qquad
V_{\mathrm{RSW},\textnormal{II}}^R(z)=E_R\{\mathrm{IF}_{\psi,\textnormal{II}}^R(O_j;z)^2\}.
\end{equation*}
Independence of the two samples eliminates a cross-covariance term. If
\begin{equation*}
n,m\to\infty,\qquad \frac{n}{m}\to\kappa\in(0,\infty),
\end{equation*}
the independent-sample central limit theorem gives
\begin{equation*}
\frac{\widehat\psi_{\mathrm{RSW}}(z)-\psi(z)}
{\{V_{\mathrm{RSW},\textnormal{II}}^A(z)/n+V_{\mathrm{RSW},\textnormal{II}}^R(z)/m\}^{1/2}}
\rightsquigarrow N(0,1).
\end{equation*}
Let
\begin{equation*}
\widehat V_{\mathrm{RSW},\textnormal{II}}^A(z)
=\mathbb P_n^A[\{\widehat{\mathrm{IF}}_{\psi,\textnormal{II},i}^A(z)\}^2],\qquad
\widehat V_{\mathrm{RSW},\textnormal{II}}^R(z)
=\mathbb P_m^R[\{\widehat{\mathrm{IF}}_{\psi,\textnormal{II},j}^R(z)\}^2].
\end{equation*}
The sample analogues of the two influence functions are mean-square consistent under the stated regularity conditions. A consistent variance estimator is therefore
\begin{equation}
\widehat{\mathrm{Var}}\{\widehat\psi_{\mathrm{RSW}}(z)\}
=\frac{\widehat V_{\mathrm{RSW},\textnormal{II}}^A(z)}{n}
+\frac{\widehat V_{\mathrm{RSW},\textnormal{II}}^R(z)}{m}.
\label{eq:supp_two_sample_variance}
\end{equation}
If the externally supplied membership-score model is treated as fixed rather than estimated in a reference sample, \(\mathrm{IF}_{\psi,\textnormal{II}}^R\), \(V_{\mathrm{RSW},\textnormal{II}}^R(z)\), and the second variance component are all omitted.
In the All\textit{of}Us application, the analysis-sample moments in Equations~\eqref{eq:if_beta_analysis}--\eqref{eq:if_psi_analysis} were IPCW-weighted, while the reference score was fitted from baseline \((V,Z,H)\) without outcome-ascertainment weights. The reference contribution in \eqref{eq:supp_two_sample_variance} used the logistic-regression sandwich covariance for \(\widehat\alpha\). The cross-fitted IPCW estimates were treated as plug-in quantities.

\subsection{Derivation of the efficiency comparison}
\label{app:oracle_efficiency}
Fix \(z\), write \(G_{i,z}=\mathbb I(Z_i=z)\), \(\Delta_i(z)=\tau(z,W_i)-\psi(z)\), and let \(q_z=E(G_{i,z}H_i)=E(G_{i,z}W_i)\). The risk-weighted representation implies \(E\{G_{i,z}W_i\Delta_i(z)\}=0\). Because \(E(H_i\mid W_i,Z_i)=W_i\), it also follows that \(E\{G_{i,z}H_i\Delta_i(z)\}=0\). Therefore,
\begin{equation*}
\phi_{H,i}(z)=q_z^{-1}G_{i,z}H_i\Delta_i(z),\qquad
\phi_{W,i}(z)=q_z^{-1}G_{i,z}W_i\Delta_i(z),
\end{equation*}
and
\begin{align*}
\mathrm{Var}\{\phi_{H,i}(z)\}
&=q_z^{-2}E\{G_{i,z}W_i\Delta_i(z)^2\},\\
\mathrm{Var}\{\phi_{W,i}(z)\}
&=q_z^{-2}E\{G_{i,z}W_i^2\Delta_i(z)^2\}.
\end{align*}
Subtracting establishes
\begin{equation*}
\mathrm{Var}\{\phi_{H,i}(z)\}-\mathrm{Var}\{\phi_{W,i}(z)\}
=q_z^{-2}E\{G_{i,z}W_i(1-W_i)\Delta_i(z)^2\}\geq0.
\end{equation*}
The inequality is strict exactly when the integrand is positive with positive probability.

Additionally, suppose \(\tau(Z_i,W_i)=b(Z_i,W_i)^\top\beta_0\) is correctly specified. Define \(b_i=b(Z_i,W_i)\), \(\varepsilon_i=\Gamma_i(\eta_0)-\tau(Z_i,W_i)\), \(Q=E(b_ib_i^\top)\), and \(c_z=E(G_{i,z}W_i b_i)\), and suppose that \(Q\) is nonsingular and
\begin{equation*}
E(\varepsilon_i\mid H_i,W_i,Z_i)=0,\qquad
\mathrm{Var}(\varepsilon_i\mid H_i,W_i,Z_i)=\sigma_\varepsilon^2.
\end{equation*}
Let \(\widehat\psi_{\mathrm{RSW}}^{*}(z)\) denote the complete RSW estimator with \(W\) treated as known but \(\beta_0\) still estimated from the orthogonal signal. The influence functions of FS-HR and this known-score RSW estimator are
\begin{align*}
\mathrm{IF}_{\mathrm{FS\text{-}HR},i}(z)
&=q_z^{-1}G_{i,z}H_i\{\Delta_i(z)+\varepsilon_i\},\\
\mathrm{IF}_{\mathrm{RSW},i}^{*}(z)
&=q_z^{-1}\{G_{i,z}W_i\Delta_i(z)+c_z^\top Q^{-1}b_i\varepsilon_i\}.
\end{align*}
The second expression follows because the full-sample least-squares projection has influence function \(Q^{-1}b_i\varepsilon_i\), and the derivative of the final aggregation with respect to \(\beta_0\) is \(c_z^\top/q_z\). Under the conditional mean-zero and homoskedasticity assumptions on \(\varepsilon_i\), all cross terms vanish and
\begin{align*}
V_{\mathrm{FS\text{-}HR}}(z)
&=q_z^{-2}\left[E\{G_{i,z}W_i\Delta_i(z)^2\}+\sigma_\varepsilon^2q_z\right],\\
V_{\mathrm{RSW}}^{*}(z)
&=q_z^{-2}\left[E\{G_{i,z}W_i^2\Delta_i(z)^2\}
+\sigma_\varepsilon^2c_z^\top Q^{-1}c_z\right].
\end{align*}
Finally, \(c_z^\top Q^{-1}b_i\) is the linear projection of \(G_{i,z}W_i\) onto the span of \(b_i\), so
\begin{equation*}
c_z^\top Q^{-1}c_z\leq E(G_{i,z}W_i^2)\leq E(G_{i,z}W_i)=q_z.
\end{equation*}
Subtracting the two variances gives the efficiency comparison displayed in the main text.

\clearpage
\section{Detailed Monte Carlo summaries for the main simulation}
\label{sec:supp-sim-tables}
Supplementary Tables~\ref{tab:supp-continuous-scenarios} and~\ref{tab:supp-binary-scenarios} report the subgroup-specific Monte Carlo summaries for the eight Setting~I scenarios. Supplementary Tables~\ref{tab:supp-continuous-scenarios-0820-m-n4} and~\ref{tab:supp-continuous-scenarios-0820-m-2n} report the corresponding continuous-outcome results for Setting~II with \(m=n/4\) and \(m=2n\), respectively. Supplementary Tables~\ref{tab:supp-binary-scenarios-0820-m-n4} and~\ref{tab:supp-binary-scenarios-0820-m-2n} report the corresponding binary-outcome results.

\begin{landscape}
\begin{table}[p]
\centering
\caption{Finite-sample Monte Carlo performance under Setting~I, with $W$ estimated internally in the analysis sample, for the RSW, FS-HR, and HR estimators with continuous outcomes. Rows report the Monte Carlo truth, empirical mean, empirical SD, absolute bias, and empirical coverage of nominal 95\% Wald confidence intervals.}
\label{tab:supp-continuous-scenarios}
\begingroup
\small
\setlength{\tabcolsep}{1.3pt}
\renewcommand{\arraystretch}{0.72}
\begin{tabular*}{\linewidth}{@{\extracolsep{\fill}}l r r r r r r r r r r r r r @{}}
\toprule
& & \multicolumn{4}{c}{$\widehat{\psi}_{\mathrm{RSW}}$} & \multicolumn{4}{c}{$\widehat{\psi}_{\mathrm{FS\text{-}HR}}$} & \multicolumn{4}{c}{$\widehat{\psi}_{\mathrm{HR}}$} \\
\cmidrule(lr){3-6}\cmidrule(lr){7-10}\cmidrule(lr){11-14}
$Z$ & $\psi_0(Z)$ & Mean & SD & Abs. Bias & 95\% Cov. & Mean & SD & Abs. Bias & 95\% Cov. & Mean & SD & Abs. Bias & 95\% Cov. \\
\midrule
\multicolumn{14}{@{}l}{\textit{Panel A. $n=500$, linear HTE}} \\
\addlinespace[0.15em]
$1$ & 3.357 & 3.326 & 0.301 & 0.031 & 0.964 & 3.376 & 0.412 & 0.019 & 0.946 & 3.338 & 0.751 & 0.019 & 0.934 \\
$2$ & 4.357 & 4.338 & 0.385 & 0.018 & 0.946 & 4.358 & 0.569 & 0.002 & 0.948 & 4.394 & 0.883 & 0.037 & 0.952 \\
$3$ & 5.364 & 5.325 & 0.364 & 0.039 & 0.944 & 5.384 & 0.526 & 0.020 & 0.932 & 5.318 & 0.753 & 0.046 & 0.956 \\
$4$ & 6.366 & 6.345 & 0.436 & 0.020 & 0.954 & 6.356 & 0.736 & 0.010 & 0.906 & 6.361 & 1.218 & 0.005 & 0.920 \\
\addlinespace[0.10em]
\multicolumn{2}{@{}l}{Avg.} & -- & 0.371 & 0.027 & 0.952 & -- & 0.561 & 0.013 & 0.933 & -- & 0.901 & 0.027 & 0.941 \\
\midrule
\multicolumn{14}{@{}l}{\textit{Panel B. $n=500$, nonlinear HTE}} \\
\addlinespace[0.15em]
$1$ & 3.022 & 3.013 & 0.352 & 0.009 & 0.970 & 3.007 & 0.515 & 0.015 & 0.942 & 3.122 & 0.835 & 0.100 & 0.942 \\
$2$ & 3.849 & 3.810 & 0.449 & 0.038 & 0.968 & 3.767 & 0.712 & 0.082 & 0.934 & 3.892 & 1.258 & 0.043 & 0.938 \\
$3$ & 4.475 & 4.479 & 0.429 & 0.003 & 0.966 & 4.518 & 0.673 & 0.042 & 0.946 & 4.586 & 0.940 & 0.111 & 0.944 \\
$4$ & 5.017 & 5.004 & 0.516 & 0.013 & 0.968 & 5.022 & 0.849 & 0.005 & 0.938 & 5.142 & 1.216 & 0.125 & 0.940 \\
\addlinespace[0.10em]
\multicolumn{2}{@{}l}{Avg.} & -- & 0.436 & 0.016 & 0.968 & -- & 0.687 & 0.036 & 0.940 & -- & 1.062 & 0.095 & 0.941 \\
\midrule
\multicolumn{14}{@{}l}{\textit{Panel C. $n=2{,}000$, linear HTE}} \\
\addlinespace[0.15em]
$1$ & 3.357 & 3.337 & 0.148 & 0.020 & 0.938 & 3.355 & 0.205 & 0.002 & 0.934 & 3.353 & 0.232 & 0.003 & 0.924 \\
$2$ & 4.357 & 4.351 & 0.187 & 0.006 & 0.934 & 4.363 & 0.279 & 0.007 & 0.940 & 4.361 & 0.320 & 0.004 & 0.926 \\
$3$ & 5.364 & 5.354 & 0.173 & 0.010 & 0.942 & 5.353 & 0.238 & 0.011 & 0.956 & 5.349 & 0.274 & 0.015 & 0.950 \\
$4$ & 6.366 & 6.371 & 0.215 & 0.005 & 0.936 & 6.364 & 0.348 & 0.001 & 0.918 & 6.369 & 0.391 & 0.004 & 0.922 \\
\addlinespace[0.10em]
\multicolumn{2}{@{}l}{Avg.} & -- & 0.181 & 0.010 & 0.938 & -- & 0.267 & 0.005 & 0.937 & -- & 0.304 & 0.006 & 0.930 \\
\midrule
\multicolumn{14}{@{}l}{\textit{Panel D. $n=2{,}000$, nonlinear HTE}} \\
\addlinespace[0.15em]
$1$ & 3.022 & 3.003 & 0.192 & 0.019 & 0.938 & 3.003 & 0.258 & 0.019 & 0.932 & 3.017 & 0.249 & 0.005 & 0.944 \\
$2$ & 3.849 & 3.843 & 0.243 & 0.005 & 0.942 & 3.853 & 0.368 & 0.005 & 0.952 & 3.859 & 0.348 & 0.010 & 0.946 \\
$3$ & 4.475 & 4.467 & 0.217 & 0.008 & 0.938 & 4.464 & 0.308 & 0.011 & 0.948 & 4.475 & 0.297 & 0.000 & 0.956 \\
$4$ & 5.017 & 4.998 & 0.257 & 0.019 & 0.956 & 4.971 & 0.404 & 0.046 & 0.938 & 4.996 & 0.379 & 0.021 & 0.946 \\
\addlinespace[0.10em]
\multicolumn{2}{@{}l}{Avg.} & -- & 0.227 & 0.013 & 0.943 & -- & 0.335 & 0.020 & 0.942 & -- & 0.318 & 0.009 & 0.948 \\
\bottomrule
\end{tabular*}
\endgroup
\end{table}
\end{landscape}

\begin{landscape}
\begin{table}[p]
\centering
\caption{Finite-sample Monte Carlo performance under Setting~I, with $W$ estimated internally in the analysis sample, for the RSW, FS-HR, and HR estimators with binary outcomes on the risk-difference scale. Rows report the Monte Carlo truth, empirical mean, empirical SD, absolute bias, and empirical coverage of nominal 95\% Wald confidence intervals.}
\label{tab:supp-binary-scenarios}
\begingroup
\small
\setlength{\tabcolsep}{1.3pt}
\renewcommand{\arraystretch}{0.72}
\begin{tabular*}{\linewidth}{@{\extracolsep{\fill}}l r r r r r r r r r r r r r @{}}
\toprule
& & \multicolumn{4}{c}{$\widehat{\psi}_{\mathrm{RSW}}$} & \multicolumn{4}{c}{$\widehat{\psi}_{\mathrm{FS\text{-}HR}}$} & \multicolumn{4}{c}{$\widehat{\psi}_{\mathrm{HR}}$} \\
\cmidrule(lr){3-6}\cmidrule(lr){7-10}\cmidrule(lr){11-14}
$Z$ & $\psi_0(Z)$ & Mean & SD & Abs. Bias & 95\% Cov. & Mean & SD & Abs. Bias & 95\% Cov. & Mean & SD & Abs. Bias & 95\% Cov. \\
\midrule
\multicolumn{14}{@{}l}{\textit{Panel A. $n=500$, linear HTE}} \\
\addlinespace[0.15em]
$1$ & 0.192 & 0.193 & 0.104 & 0.001 & 0.940 & 0.187 & 0.142 & 0.005 & 0.932 & 0.195 & 0.172 & 0.003 & 0.952 \\
$2$ & 0.292 & 0.285 & 0.130 & 0.007 & 0.934 & 0.282 & 0.204 & 0.010 & 0.926 & 0.285 & 0.249 & 0.007 & 0.940 \\
$3$ & 0.392 & 0.386 & 0.117 & 0.006 & 0.946 & 0.392 & 0.178 & 0.000 & 0.926 & 0.388 & 0.233 & 0.004 & 0.922 \\
$4$ & 0.492 & 0.491 & 0.139 & 0.001 & 0.942 & 0.515 & 0.215 & 0.023 & 0.912 & 0.506 & 0.282 & 0.014 & 0.914 \\
\addlinespace[0.10em]
\multicolumn{2}{@{}l}{Avg.} & -- & 0.122 & 0.004 & 0.941 & -- & 0.185 & 0.009 & 0.924 & -- & 0.234 & 0.007 & 0.932 \\
\midrule
\multicolumn{14}{@{}l}{\textit{Panel B. $n=500$, nonlinear HTE}} \\
\addlinespace[0.15em]
$1$ & 0.035 & 0.035 & 0.102 & 0.000 & 0.940 & 0.034 & 0.135 & 0.002 & 0.960 & 0.039 & 0.174 & 0.004 & 0.948 \\
$2$ & 0.125 & 0.114 & 0.129 & 0.011 & 0.960 & 0.105 & 0.206 & 0.020 & 0.946 & 0.099 & 0.246 & 0.026 & 0.952 \\
$3$ & 0.275 & 0.270 & 0.121 & 0.005 & 0.950 & 0.278 & 0.187 & 0.003 & 0.938 & 0.269 & 0.240 & 0.006 & 0.944 \\
$4$ & 0.485 & 0.482 & 0.136 & 0.003 & 0.954 & 0.507 & 0.224 & 0.021 & 0.900 & 0.506 & 0.293 & 0.021 & 0.912 \\
\addlinespace[0.10em]
\multicolumn{2}{@{}l}{Avg.} & -- & 0.122 & 0.005 & 0.951 & -- & 0.188 & 0.011 & 0.936 & -- & 0.238 & 0.014 & 0.939 \\
\midrule
\multicolumn{14}{@{}l}{\textit{Panel C. $n=2{,}000$, linear HTE}} \\
\addlinespace[0.15em]
$1$ & 0.192 & 0.193 & 0.050 & 0.001 & 0.938 & 0.190 & 0.067 & 0.002 & 0.948 & 0.190 & 0.071 & 0.002 & 0.942 \\
$2$ & 0.292 & 0.288 & 0.061 & 0.004 & 0.938 & 0.288 & 0.089 & 0.004 & 0.962 & 0.287 & 0.092 & 0.005 & 0.962 \\
$3$ & 0.392 & 0.392 & 0.056 & 0.000 & 0.944 & 0.395 & 0.081 & 0.003 & 0.952 & 0.392 & 0.083 & 0.000 & 0.956 \\
$4$ & 0.492 & 0.495 & 0.058 & 0.002 & 0.966 & 0.495 & 0.094 & 0.003 & 0.950 & 0.498 & 0.098 & 0.006 & 0.956 \\
\addlinespace[0.10em]
\multicolumn{2}{@{}l}{Avg.} & -- & 0.056 & 0.002 & 0.946 & -- & 0.083 & 0.003 & 0.953 & -- & 0.086 & 0.003 & 0.954 \\
\midrule
\multicolumn{14}{@{}l}{\textit{Panel D. $n=2{,}000$, nonlinear HTE}} \\
\addlinespace[0.15em]
$1$ & 0.035 & 0.036 & 0.049 & 0.000 & 0.948 & 0.034 & 0.063 & 0.001 & 0.958 & 0.035 & 0.067 & 0.001 & 0.942 \\
$2$ & 0.125 & 0.120 & 0.060 & 0.005 & 0.950 & 0.123 & 0.093 & 0.003 & 0.954 & 0.124 & 0.094 & 0.002 & 0.960 \\
$3$ & 0.275 & 0.274 & 0.059 & 0.002 & 0.936 & 0.273 & 0.090 & 0.002 & 0.944 & 0.272 & 0.091 & 0.004 & 0.944 \\
$4$ & 0.485 & 0.488 & 0.058 & 0.002 & 0.964 & 0.488 & 0.095 & 0.002 & 0.954 & 0.491 & 0.102 & 0.005 & 0.952 \\
\addlinespace[0.10em]
\multicolumn{2}{@{}l}{Avg.} & -- & 0.056 & 0.002 & 0.950 & -- & 0.085 & 0.002 & 0.952 & -- & 0.088 & 0.003 & 0.950 \\
\bottomrule
\end{tabular*}
\endgroup
\end{table}
\end{landscape}

\begin{landscape}
\begin{table}[p]
\centering

\caption{Finite-sample Monte Carlo performance under Setting~II, with $W$ estimated in an independent reference sample of size $m=n/4$, for the RSW, FS-HR, and HR estimators with continuous outcomes. Rows report the Monte Carlo truth, empirical mean, empirical SD, absolute bias, and empirical coverage of nominal 95\% Wald confidence intervals.}
\label{tab:supp-continuous-scenarios-0820-m-n4}
\begingroup
\footnotesize
\setlength{\tabcolsep}{1.3pt}
\renewcommand{\arraystretch}{0.72}
\begin{tabular*}{\linewidth}{@{\extracolsep{\fill}}l r r r r r r r r r r r r r @{}}
\toprule
& & \multicolumn{4}{c}{$\widehat{\psi}_{\mathrm{RSW}}$} & \multicolumn{4}{c}{$\widehat{\psi}_{\mathrm{FS\text{-}HR}}$} & \multicolumn{4}{c}{$\widehat{\psi}_{\mathrm{HR}}$} \\
\cmidrule(lr){3-6}\cmidrule(lr){7-10}\cmidrule(lr){11-14}
$Z$ & $\psi_0(Z)$ & Mean & SD & Abs. Bias & 95\% Cov. & Mean & SD & Abs. Bias & 95\% Cov. & Mean & SD & Abs. Bias & 95\% Cov. \\
\midrule
\multicolumn{14}{@{}l}{\textit{Panel A. $n=500$, linear HTE}} \\
\addlinespace[0.15em]
$1$ & 3.357 & 3.292 & 0.357 & 0.064 & 0.966 & 3.368 & 0.401 & 0.011 & 0.950 & 3.338 & 0.751 & 0.019 & 0.934 \\
$2$ & 4.357 & 4.299 & 0.428 & 0.058 & 0.966 & 4.364 & 0.584 & 0.007 & 0.946 & 4.394 & 0.883 & 0.037 & 0.952 \\
$3$ & 5.364 & 5.306 & 0.410 & 0.058 & 0.946 & 5.390 & 0.530 & 0.026 & 0.932 & 5.318 & 0.753 & 0.046 & 0.956 \\
$4$ & 6.366 & 6.321 & 0.470 & 0.045 & 0.966 & 6.365 & 0.762 & 0.000 & 0.904 & 6.361 & 1.218 & 0.005 & 0.920 \\
\addlinespace[0.10em]
\multicolumn{2}{@{}l}{Avg.} & -- & 0.416 & 0.056 & 0.961 & -- & 0.569 & 0.011 & 0.933 & -- & 0.901 & 0.027 & 0.940 \\
\midrule
\multicolumn{14}{@{}l}{\textit{Panel B. $n=500$, nonlinear HTE}} \\
\addlinespace[0.15em]
$1$ & 3.022 & 2.998 & 0.393 & 0.024 & 0.958 & 3.006 & 0.514 & 0.016 & 0.942 & 3.122 & 0.835 & 0.100 & 0.942 \\
$2$ & 3.849 & 3.792 & 0.480 & 0.057 & 0.964 & 3.768 & 0.711 & 0.080 & 0.928 & 3.892 & 1.258 & 0.043 & 0.938 \\
$3$ & 4.475 & 4.454 & 0.439 & 0.022 & 0.968 & 4.521 & 0.675 & 0.046 & 0.946 & 4.586 & 0.940 & 0.111 & 0.944 \\
$4$ & 5.017 & 4.987 & 0.545 & 0.030 & 0.968 & 5.019 & 0.848 & 0.002 & 0.934 & 5.142 & 1.216 & 0.125 & 0.940 \\
\addlinespace[0.10em]
\multicolumn{2}{@{}l}{Avg.} & -- & 0.464 & 0.033 & 0.964 & -- & 0.687 & 0.036 & 0.938 & -- & 1.062 & 0.095 & 0.941 \\
\midrule
\multicolumn{14}{@{}l}{\textit{Panel C. $n=2{,}000$, linear HTE}} \\
\addlinespace[0.15em]
$1$ & 3.357 & 3.322 & 0.177 & 0.034 & 0.958 & 3.357 & 0.206 & 0.001 & 0.936 & 3.353 & 0.232 & 0.003 & 0.924 \\
$2$ & 4.357 & 4.338 & 0.215 & 0.019 & 0.958 & 4.365 & 0.289 & 0.009 & 0.932 & 4.361 & 0.320 & 0.004 & 0.926 \\
$3$ & 5.364 & 5.353 & 0.197 & 0.012 & 0.952 & 5.358 & 0.240 & 0.007 & 0.954 & 5.349 & 0.274 & 0.015 & 0.950 \\
$4$ & 6.366 & 6.364 & 0.234 & 0.002 & 0.954 & 6.361 & 0.350 & 0.004 & 0.930 & 6.369 & 0.391 & 0.004 & 0.922 \\
\addlinespace[0.10em]
\multicolumn{2}{@{}l}{Avg.} & -- & 0.206 & 0.017 & 0.956 & -- & 0.271 & 0.005 & 0.938 & -- & 0.304 & 0.006 & 0.930 \\
\midrule
\multicolumn{14}{@{}l}{\textit{Panel D. $n=2{,}000$, nonlinear HTE}} \\
\addlinespace[0.15em]
$1$ & 3.022 & 3.002 & 0.203 & 0.020 & 0.944 & 3.003 & 0.258 & 0.019 & 0.932 & 3.017 & 0.249 & 0.005 & 0.944 \\
$2$ & 3.849 & 3.839 & 0.256 & 0.010 & 0.946 & 3.853 & 0.368 & 0.005 & 0.950 & 3.859 & 0.348 & 0.010 & 0.946 \\
$3$ & 4.475 & 4.466 & 0.230 & 0.010 & 0.948 & 4.464 & 0.308 & 0.011 & 0.948 & 4.475 & 0.297 & 0.000 & 0.956 \\
$4$ & 5.017 & 4.999 & 0.262 & 0.018 & 0.966 & 4.970 & 0.403 & 0.046 & 0.942 & 4.996 & 0.379 & 0.021 & 0.946 \\
\addlinespace[0.10em]
\multicolumn{2}{@{}l}{Avg.} & -- & 0.238 & 0.014 & 0.951 & -- & 0.334 & 0.020 & 0.943 & -- & 0.318 & 0.009 & 0.948 \\
\bottomrule
\end{tabular*}
\endgroup
\end{table}
\end{landscape}

\begin{landscape}
\begin{table}[p]
\centering

\caption{Finite-sample Monte Carlo performance under Setting~II, with $W$ estimated in an independent reference sample of size $m=2n$, for the RSW, FS-HR, and HR estimators with continuous outcomes. Rows report the Monte Carlo truth, empirical mean, empirical SD, absolute bias, and empirical coverage of nominal 95\% Wald confidence intervals.}
\label{tab:supp-continuous-scenarios-0820-m-2n}
\begingroup
\footnotesize
\setlength{\tabcolsep}{1.3pt}
\renewcommand{\arraystretch}{0.72}
\begin{tabular*}{\linewidth}{@{\extracolsep{\fill}}l r r r r r r r r r r r r r @{}}
\toprule
& & \multicolumn{4}{c}{$\widehat{\psi}_{\mathrm{RSW}}$} & \multicolumn{4}{c}{$\widehat{\psi}_{\mathrm{FS\text{-}HR}}$} & \multicolumn{4}{c}{$\widehat{\psi}_{\mathrm{HR}}$} \\
\cmidrule(lr){3-6}\cmidrule(lr){7-10}\cmidrule(lr){11-14}
$Z$ & $\psi_0(Z)$ & Mean & SD & Abs. Bias & 95\% Cov. & Mean & SD & Abs. Bias & 95\% Cov. & Mean & SD & Abs. Bias & 95\% Cov. \\
\midrule
\multicolumn{14}{@{}l}{\textit{Panel A. $n=500$, linear HTE}} \\
\addlinespace[0.15em]
$1$ & 3.357 & 3.351 & 0.276 & 0.006 & 0.960 & 3.370 & 0.397 & 0.013 & 0.956 & 3.338 & 0.751 & 0.019 & 0.934 \\
$2$ & 4.357 & 4.346 & 0.363 & 0.011 & 0.944 & 4.363 & 0.546 & 0.006 & 0.958 & 4.394 & 0.883 & 0.037 & 0.952 \\
$3$ & 5.364 & 5.334 & 0.348 & 0.030 & 0.942 & 5.372 & 0.507 & 0.008 & 0.942 & 5.318 & 0.753 & 0.046 & 0.956 \\
$4$ & 6.366 & 6.349 & 0.407 & 0.017 & 0.962 & 6.369 & 0.708 & 0.003 & 0.904 & 6.361 & 1.218 & 0.005 & 0.920 \\
\addlinespace[0.10em]
\multicolumn{2}{@{}l}{Avg.} & -- & 0.348 & 0.016 & 0.952 & -- & 0.540 & 0.007 & 0.940 & -- & 0.901 & 0.027 & 0.940 \\
\midrule
\multicolumn{14}{@{}l}{\textit{Panel B. $n=500$, nonlinear HTE}} \\
\addlinespace[0.15em]
$1$ & 3.022 & 3.017 & 0.359 & 0.005 & 0.962 & 3.006 & 0.516 & 0.016 & 0.942 & 3.122 & 0.835 & 0.100 & 0.942 \\
$2$ & 3.849 & 3.821 & 0.458 & 0.028 & 0.970 & 3.764 & 0.710 & 0.085 & 0.936 & 3.892 & 1.258 & 0.043 & 0.938 \\
$3$ & 4.475 & 4.477 & 0.433 & 0.001 & 0.958 & 4.520 & 0.669 & 0.044 & 0.950 & 4.586 & 0.940 & 0.111 & 0.944 \\
$4$ & 5.017 & 5.010 & 0.545 & 0.007 & 0.956 & 5.017 & 0.849 & 0.001 & 0.934 & 5.142 & 1.216 & 0.125 & 0.940 \\
\addlinespace[0.10em]
\multicolumn{2}{@{}l}{Avg.} & -- & 0.449 & 0.010 & 0.961 & -- & 0.686 & 0.037 & 0.941 & -- & 1.062 & 0.095 & 0.941 \\
\midrule
\multicolumn{14}{@{}l}{\textit{Panel C. $n=2{,}000$, linear HTE}} \\
\addlinespace[0.15em]
$1$ & 3.357 & 3.350 & 0.139 & 0.007 & 0.954 & 3.356 & 0.205 & 0.001 & 0.928 & 3.353 & 0.232 & 0.003 & 0.924 \\
$2$ & 4.357 & 4.359 & 0.179 & 0.002 & 0.948 & 4.364 & 0.275 & 0.007 & 0.938 & 4.361 & 0.320 & 0.004 & 0.926 \\
$3$ & 5.364 & 5.355 & 0.166 & 0.009 & 0.942 & 5.354 & 0.232 & 0.010 & 0.958 & 5.349 & 0.274 & 0.015 & 0.950 \\
$4$ & 6.366 & 6.367 & 0.212 & 0.002 & 0.942 & 6.362 & 0.342 & 0.003 & 0.922 & 6.369 & 0.391 & 0.004 & 0.922 \\
\addlinespace[0.10em]
\multicolumn{2}{@{}l}{Avg.} & -- & 0.174 & 0.005 & 0.946 & -- & 0.264 & 0.005 & 0.936 & -- & 0.304 & 0.006 & 0.930 \\
\midrule
\multicolumn{14}{@{}l}{\textit{Panel D. $n=2{,}000$, nonlinear HTE}} \\
\addlinespace[0.15em]
$1$ & 3.022 & 3.003 & 0.190 & 0.019 & 0.942 & 3.003 & 0.258 & 0.019 & 0.932 & 3.017 & 0.249 & 0.005 & 0.944 \\
$2$ & 3.849 & 3.840 & 0.247 & 0.009 & 0.930 & 3.854 & 0.368 & 0.005 & 0.952 & 3.859 & 0.348 & 0.010 & 0.946 \\
$3$ & 4.475 & 4.466 & 0.217 & 0.009 & 0.936 & 4.463 & 0.308 & 0.012 & 0.950 & 4.475 & 0.297 & 0.000 & 0.956 \\
$4$ & 5.017 & 4.991 & 0.255 & 0.026 & 0.956 & 4.971 & 0.404 & 0.046 & 0.940 & 4.996 & 0.379 & 0.021 & 0.946 \\
\addlinespace[0.10em]
\multicolumn{2}{@{}l}{Avg.} & -- & 0.227 & 0.016 & 0.941 & -- & 0.335 & 0.021 & 0.943 & -- & 0.318 & 0.009 & 0.948 \\
\bottomrule
\end{tabular*}
\endgroup
\end{table}
\end{landscape}

\begin{landscape}
\begin{table}[p]
\centering

\caption{Finite-sample Monte Carlo performance under Setting~II, with $W$ estimated in an independent reference sample of size $m=n/4$, for the RSW, FS-HR, and HR estimators with binary outcomes on the risk-difference scale. Rows report the Monte Carlo truth, empirical mean, empirical SD, absolute bias, and empirical coverage of nominal 95\% Wald confidence intervals.}
\label{tab:supp-binary-scenarios-0820-m-n4}
\begingroup
\footnotesize
\setlength{\tabcolsep}{1.3pt}
\renewcommand{\arraystretch}{0.72}
\begin{tabular*}{\linewidth}{@{\extracolsep{\fill}}l r r r r r r r r r r r r r @{}}
\toprule
& & \multicolumn{4}{c}{$\widehat{\psi}_{\mathrm{RSW}}$} & \multicolumn{4}{c}{$\widehat{\psi}_{\mathrm{FS\text{-}HR}}$} & \multicolumn{4}{c}{$\widehat{\psi}_{\mathrm{HR}}$} \\
\cmidrule(lr){3-6}\cmidrule(lr){7-10}\cmidrule(lr){11-14}
$Z$ & $\psi_0(Z)$ & Mean & SD & Abs. Bias & 95\% Cov. & Mean & SD & Abs. Bias & 95\% Cov. & Mean & SD & Abs. Bias & 95\% Cov. \\
\midrule
\multicolumn{14}{@{}l}{\textit{Panel A. $n=500$, linear HTE}} \\
\addlinespace[0.15em]
$1$ & 0.192 & 0.192 & 0.106 & 0.000 & 0.942 & 0.187 & 0.142 & 0.005 & 0.934 & 0.195 & 0.172 & 0.003 & 0.952 \\
$2$ & 0.292 & 0.285 & 0.129 & 0.007 & 0.940 & 0.282 & 0.204 & 0.010 & 0.926 & 0.285 & 0.249 & 0.007 & 0.940 \\
$3$ & 0.392 & 0.385 & 0.118 & 0.007 & 0.946 & 0.392 & 0.177 & 0.000 & 0.924 & 0.388 & 0.233 & 0.004 & 0.922 \\
$4$ & 0.492 & 0.491 & 0.139 & 0.001 & 0.944 & 0.516 & 0.214 & 0.024 & 0.910 & 0.506 & 0.282 & 0.014 & 0.914 \\
\addlinespace[0.10em]
\multicolumn{2}{@{}l}{Avg.} & -- & 0.123 & 0.004 & 0.943 & -- & 0.184 & 0.010 & 0.924 & -- & 0.234 & 0.007 & 0.932 \\
\midrule
\multicolumn{14}{@{}l}{\textit{Panel B. $n=500$, nonlinear HTE}} \\
\addlinespace[0.15em]
$1$ & 0.035 & 0.036 & 0.104 & 0.000 & 0.942 & 0.033 & 0.135 & 0.002 & 0.960 & 0.039 & 0.174 & 0.004 & 0.948 \\
$2$ & 0.125 & 0.114 & 0.131 & 0.012 & 0.960 & 0.105 & 0.206 & 0.020 & 0.948 & 0.099 & 0.246 & 0.026 & 0.952 \\
$3$ & 0.275 & 0.269 & 0.123 & 0.006 & 0.952 & 0.278 & 0.186 & 0.003 & 0.938 & 0.269 & 0.240 & 0.006 & 0.944 \\
$4$ & 0.485 & 0.483 & 0.136 & 0.002 & 0.962 & 0.507 & 0.223 & 0.022 & 0.904 & 0.506 & 0.293 & 0.021 & 0.912 \\
\addlinespace[0.10em]
\multicolumn{2}{@{}l}{Avg.} & -- & 0.123 & 0.005 & 0.954 & -- & 0.187 & 0.012 & 0.938 & -- & 0.238 & 0.014 & 0.939 \\
\midrule
\multicolumn{14}{@{}l}{\textit{Panel C. $n=2{,}000$, linear HTE}} \\
\addlinespace[0.15em]
$1$ & 0.192 & 0.192 & 0.050 & 0.000 & 0.936 & 0.190 & 0.067 & 0.002 & 0.950 & 0.190 & 0.071 & 0.002 & 0.942 \\
$2$ & 0.292 & 0.288 & 0.061 & 0.004 & 0.936 & 0.288 & 0.089 & 0.004 & 0.964 & 0.287 & 0.092 & 0.005 & 0.962 \\
$3$ & 0.392 & 0.392 & 0.057 & 0.000 & 0.946 & 0.395 & 0.081 & 0.003 & 0.952 & 0.392 & 0.083 & 0.000 & 0.956 \\
$4$ & 0.492 & 0.495 & 0.058 & 0.003 & 0.964 & 0.494 & 0.094 & 0.002 & 0.950 & 0.498 & 0.098 & 0.006 & 0.956 \\
\addlinespace[0.10em]
\multicolumn{2}{@{}l}{Avg.} & -- & 0.057 & 0.002 & 0.946 & -- & 0.083 & 0.003 & 0.954 & -- & 0.086 & 0.003 & 0.954 \\
\midrule
\multicolumn{14}{@{}l}{\textit{Panel D. $n=2{,}000$, nonlinear HTE}} \\
\addlinespace[0.15em]
$1$ & 0.035 & 0.036 & 0.049 & 0.000 & 0.948 & 0.034 & 0.063 & 0.001 & 0.958 & 0.035 & 0.067 & 0.001 & 0.942 \\
$2$ & 0.125 & 0.120 & 0.060 & 0.005 & 0.948 & 0.123 & 0.093 & 0.003 & 0.956 & 0.124 & 0.094 & 0.002 & 0.960 \\
$3$ & 0.275 & 0.274 & 0.059 & 0.002 & 0.930 & 0.274 & 0.090 & 0.002 & 0.944 & 0.272 & 0.091 & 0.004 & 0.944 \\
$4$ & 0.485 & 0.487 & 0.058 & 0.002 & 0.962 & 0.488 & 0.095 & 0.002 & 0.954 & 0.491 & 0.102 & 0.005 & 0.952 \\
\addlinespace[0.10em]
\multicolumn{2}{@{}l}{Avg.} & -- & 0.056 & 0.002 & 0.947 & -- & 0.085 & 0.002 & 0.953 & -- & 0.088 & 0.003 & 0.950 \\
\bottomrule
\end{tabular*}
\endgroup
\end{table}
\end{landscape}

\begin{landscape}
\begin{table}[p]
\centering

\caption{Finite-sample Monte Carlo performance under Setting~II, with $W$ estimated in an independent reference sample of size $m=2n$, for the RSW, FS-HR, and HR estimators with binary outcomes on the risk-difference scale. Rows report the Monte Carlo truth, empirical mean, empirical SD, absolute bias, and empirical coverage of nominal 95\% Wald confidence intervals.}
\label{tab:supp-binary-scenarios-0820-m-2n}
\begingroup
\footnotesize
\setlength{\tabcolsep}{1.3pt}
\renewcommand{\arraystretch}{0.72}
\begin{tabular*}{\linewidth}{@{\extracolsep{\fill}}l r r r r r r r r r r r r r @{}}
\toprule
& & \multicolumn{4}{c}{$\widehat{\psi}_{\mathrm{RSW}}$} & \multicolumn{4}{c}{$\widehat{\psi}_{\mathrm{FS\text{-}HR}}$} & \multicolumn{4}{c}{$\widehat{\psi}_{\mathrm{HR}}$} \\
\cmidrule(lr){3-6}\cmidrule(lr){7-10}\cmidrule(lr){11-14}
$Z$ & $\psi_0(Z)$ & Mean & SD & Abs. Bias & 95\% Cov. & Mean & SD & Abs. Bias & 95\% Cov. & Mean & SD & Abs. Bias & 95\% Cov. \\
\midrule
\multicolumn{14}{@{}l}{\textit{Panel A. $n=500$, linear HTE}} \\
\addlinespace[0.15em]
$1$ & 0.192 & 0.193 & 0.104 & 0.001 & 0.942 & 0.187 & 0.142 & 0.005 & 0.934 & 0.195 & 0.172 & 0.003 & 0.952 \\
$2$ & 0.292 & 0.285 & 0.129 & 0.007 & 0.938 & 0.283 & 0.204 & 0.009 & 0.928 & 0.285 & 0.249 & 0.007 & 0.940 \\
$3$ & 0.392 & 0.386 & 0.117 & 0.006 & 0.944 & 0.392 & 0.177 & 0.000 & 0.928 & 0.388 & 0.233 & 0.004 & 0.922 \\
$4$ & 0.492 & 0.491 & 0.137 & 0.001 & 0.944 & 0.515 & 0.213 & 0.023 & 0.906 & 0.506 & 0.282 & 0.014 & 0.914 \\
\addlinespace[0.10em]
\multicolumn{2}{@{}l}{Avg.} & -- & 0.122 & 0.004 & 0.942 & -- & 0.184 & 0.009 & 0.924 & -- & 0.234 & 0.007 & 0.932 \\
\midrule
\multicolumn{14}{@{}l}{\textit{Panel B. $n=500$, nonlinear HTE}} \\
\addlinespace[0.15em]
$1$ & 0.035 & 0.036 & 0.102 & 0.000 & 0.940 & 0.034 & 0.134 & 0.001 & 0.960 & 0.039 & 0.174 & 0.004 & 0.948 \\
$2$ & 0.125 & 0.115 & 0.129 & 0.011 & 0.958 & 0.105 & 0.205 & 0.020 & 0.948 & 0.099 & 0.246 & 0.026 & 0.952 \\
$3$ & 0.275 & 0.270 & 0.121 & 0.005 & 0.950 & 0.278 & 0.186 & 0.003 & 0.942 & 0.269 & 0.240 & 0.006 & 0.944 \\
$4$ & 0.485 & 0.484 & 0.135 & 0.001 & 0.956 & 0.507 & 0.222 & 0.021 & 0.904 & 0.506 & 0.293 & 0.021 & 0.912 \\
\addlinespace[0.10em]
\multicolumn{2}{@{}l}{Avg.} & -- & 0.122 & 0.004 & 0.951 & -- & 0.187 & 0.011 & 0.938 & -- & 0.238 & 0.014 & 0.939 \\
\midrule
\multicolumn{14}{@{}l}{\textit{Panel C. $n=2{,}000$, linear HTE}} \\
\addlinespace[0.15em]
$1$ & 0.192 & 0.193 & 0.050 & 0.001 & 0.932 & 0.190 & 0.067 & 0.002 & 0.948 & 0.190 & 0.071 & 0.002 & 0.942 \\
$2$ & 0.292 & 0.288 & 0.061 & 0.004 & 0.934 & 0.288 & 0.089 & 0.004 & 0.962 & 0.287 & 0.092 & 0.005 & 0.962 \\
$3$ & 0.392 & 0.392 & 0.057 & 0.000 & 0.944 & 0.395 & 0.081 & 0.003 & 0.952 & 0.392 & 0.083 & 0.000 & 0.956 \\
$4$ & 0.492 & 0.495 & 0.058 & 0.003 & 0.966 & 0.495 & 0.094 & 0.002 & 0.950 & 0.498 & 0.098 & 0.006 & 0.956 \\
\addlinespace[0.10em]
\multicolumn{2}{@{}l}{Avg.} & -- & 0.057 & 0.002 & 0.944 & -- & 0.083 & 0.003 & 0.953 & -- & 0.086 & 0.003 & 0.954 \\
\midrule
\multicolumn{14}{@{}l}{\textit{Panel D. $n=2{,}000$, nonlinear HTE}} \\
\addlinespace[0.15em]
$1$ & 0.035 & 0.036 & 0.049 & 0.000 & 0.944 & 0.034 & 0.063 & 0.001 & 0.958 & 0.035 & 0.067 & 0.001 & 0.942 \\
$2$ & 0.125 & 0.120 & 0.060 & 0.005 & 0.946 & 0.123 & 0.093 & 0.003 & 0.954 & 0.124 & 0.094 & 0.002 & 0.960 \\
$3$ & 0.275 & 0.274 & 0.059 & 0.002 & 0.938 & 0.273 & 0.090 & 0.002 & 0.944 & 0.272 & 0.091 & 0.004 & 0.944 \\
$4$ & 0.485 & 0.488 & 0.058 & 0.002 & 0.966 & 0.488 & 0.095 & 0.002 & 0.954 & 0.491 & 0.102 & 0.005 & 0.952 \\
\addlinespace[0.10em]
\multicolumn{2}{@{}l}{Avg.} & -- & 0.056 & 0.002 & 0.948 & -- & 0.085 & 0.002 & 0.952 & -- & 0.088 & 0.003 & 0.950 \\
\bottomrule
\end{tabular*}
\endgroup
\end{table}
\end{landscape}

\bibliographystyle{plainnat}
\bibliography{latex/supplementary_refs}